\documentclass{aa}

\usepackage{graphicx}
\usepackage{xcolor}
\usepackage{multirow}
\usepackage{txfonts}

\definecolor{myblue}{rgb}{0.00, 0.0, 0.9}
\definecolor{myred}{rgb}{0.90, 0.0, 0.0}
\definecolor{mygreen}{rgb}{0.0, 0.7, 0.0}

\usepackage[breaklinks,  citecolor=myblue, linkcolor=myred, urlcolor=purple, colorlinks=true, debug, baseurl=' ']{hyperref}

\newcommand{\hi}{H{\sc i}~}
\newcommand{\HI}{H{\sc i}}

\begin{document} 

\title{Evidence for Line-of-Sight Frequency Decorrelation \\ of Polarized Dust Emission in {\it Planck} Data}

\titlerunning{LOS frequency decorrelation of dust polarization in {\it Planck}}

   \author{V.~Pelgrims \inst{1} \fnmsep \inst{2} \fnmsep \thanks{pelgrims@physics.uoc.gr}, S.~E.~Clark \inst{3}, B.~S.~Hensley \inst{4}, G.~V.~Panopoulou \inst{5}, V.~Pavlidou \inst{1} \fnmsep \inst{2}, K.~Tassis \inst{1} \fnmsep \inst{2}, H.K.~Eriksen \inst{6}, I.~K.~Wehus \inst{6}  }
\institute{Institute of Astrophysics, Foundation for Research and Technology-Hellas, GR-71110 Heraklion, Greece
\and
Department of Physics, and Institute for Theoretical and Computational Physics, University of Crete, GR-70013 Heraklion, Greece
\and Institute for Advanced Study, 1 Einstein Drive, Princeton, NJ 08540, USA
\and Spitzer Fellow, Department of Astrophysical Sciences, Princeton University, Princeton, NJ 08544, USA
\and Hubble Fellow, California Institute of Technology, MC350-17, 1200 East California Boulevard, Pasadena, CA 91125, USA
\and Institute of Theoretical Astrophysics, University of Oslo, PO Box 1029 Blindern, 0315 Oslo, Norway
}
\authorrunning{Pelgrims et al.}
   \date{Received December 23, 2020; accepted January 19, 2021}

  \abstract{
If a single line of sight (LOS) intercepts multiple dust clouds having different spectral energy distributions (SEDs) and magnetic field orientations, then the frequency scaling of each of the Stokes $Q$ and $U$ parameters of the thermal dust emission may be different, a phenomenon we refer to as LOS frequency decorrelation.
We present first evidence for LOS frequency decorrelation in \textit{Planck} data by using independent measurements of neutral hydrogen (\HI) emission to probe the 3D structure of the magnetized ISM. We use \HI-based measurements of the number of clouds per LOS and the magnetic field orientation in each cloud to select two sets of sightlines: 
  (i) a target sample of pixels that are likely to exhibit LOS frequency decorrelation and (ii) a control sample of pixels that lack complex LOS structure.
  We test the null hypothesis that LOS frequency decorrelation is not detectable in {\it Planck} 353 and 217~GHz polarization data at high Galactic latitudes.
  We find that the data reject the null hypothesis at high significance, showing that the combined effect of polarization angle variation with frequency and depolarization are detected in the target sample. This detection is robust against choice of CMB map and map-making pipeline. The observed change in polarization angle due to LOS frequency decorrelation is detectable above the {\it Planck} noise level. The probability that the detected effect is due to noise alone ranges from $5\times 10^{-2}$ to $4\times 10^{-7}$, depending on the CMB subtraction algorithm and treatment of residual systematics;
  correcting for residual systematics consistently increases the significance of the effect.
  Within the target sample, the LOS decorrelation effect is stronger for sightlines with more misaligned magnetic fields, as expected. 
  With our sample, we estimate that an intrinsic variation of $\sim15\%$ in the ratio of 353 to 217~GHz polarized emission between clouds is sufficient to reproduce the measured effect.
  Our finding underlines the importance of ongoing studies to map the three-dimensional structure of the magnetized and dusty ISM that could ultimately help component separation methods to account for frequency decorrelation effects in CMB polarization studies.
}
   \keywords{
   ISM: dust, magnetic fields --
   submillimeter: ISM --
   (cosmology) cosmic background radiation --
   inflation --
   polarization
   	               }
   	               
   \maketitle
%

\section{Introduction}
\label{sec:intro}

Cosmic Microwave Background (CMB) polarization experiments have reached a sensitivity sufficient to demonstrate that, even in the most diffuse regions of the sky, cosmological signals of interest lie below the polarized emission from Galactic foregrounds \citep{B2K2018,PlaIV2018}. In particular, the B-mode signature from primordial gravitational waves \citep{2016ARA&A..54..227K}, quantified by the tensor-to-scalar ratio $r$, is now constrained to be at least $\sim$ten times fainter than B-mode emission from Galactic dust at 150\,GHz, even in the diffuse BICEP/\textit{Keck} region \citep{B2K2018}. Next-generation experiments like the Simons Observatory \citep{Ade2019}, CMB-S4 \citep{Aba2016}, and LiteBIRD \citep{Suz2018} seek constraints on $r$ that improve on current upper limits by an order of magnitude or more and will thus require foreground mitigation to the percent level or better.

One of the most challenging aspect of modeling dust foregrounds is that the spectral energy distribution (SED) of dust emission is not uniform across the sky. Variations in dust temperature and opacity law are now well-attested across the Galaxy (e.g., \citealt{FDS99}; \citealt{PlaXI2014}; \citealt{Mei15}; \citealt{PlaIV2018}; \citealt{Irf19}), with evidence for correlations with gas velocity (\citealt{PlaXXIV2011}; \citealt{PlaXI2014}), strength of the ambient radiation field \citep{PlaXXIX,Fan15}, and location in the Galactic disk \citep{Sch16}. 

Such variations greatly restrict the ability to use maps of dust emission at one frequency to constrain dust emission at another frequency--i.e., two maps at different frequencies differ by more than just an overall multiplicative factor ({\it frequency decorrelation}). The three-dimensional (3D) structure of the interstellar medium adds to the complexity of this problem \citep{Tas2015}. If a single line of sight (LOS) intercepts multiple dust clouds having different SEDs and magnetic field orientations, then the frequency scaling of each of the Stokes $Q$ and $U$ parameters may be different even in a single pixel (LOS frequency decorrelation). Frequency decorrelation has already been identified as a critical uncertainty in current $r$ constraints and will be even more acute at higher sensitivities \citep{B2K2018,CMBS420}.

Frequency decorrelation is often quantified at the power spectrum level through the ratio $R_\ell^{BB}$ of the $BB$ cross-spectrum of two frequencies at some multipole $\ell$ to the geometric mean of their auto-spectra \citep{PlanckL2017}. Computing $R_\ell^{BB}$ over large areas of the {\it Planck} polarization maps at 353 and 217\,GHz, the channels with the greatest sensitivity to polarized dust emission, has yielded only limits of $R_\ell^{BB} \gtrsim 0.98$ \citep{ShSl2018,PlaXI2018}. While this limit suggests frequency decorrelation may not be a limiting concern if $r \gtrsim 0.01$, \citet{PlaXI2018} caution that the level of decorrelation may be variable across the sky with some limited sky regions potentially having much greater values.

LOS frequency decorrelation can have a particularly pernicious effect on parametric component separation methods working at the map level, especially if the SEDs of Stokes $Q$ and $U$ are not modeled with independent parameters \citep{PD2017, Tuhin2017, PFB2017, Hen2018, MKD2018, CMBS420}. New techniques employing moment decomposition \citep{Chl2017} have shown promise for mitigating LOS averaging of dust SEDs in polarization at the expense of additional parameters \citep{Man2019,Rem2020}. Distortions of the SED from effects like LOS frequency decorrelation are also important for power spectrum-based modeling of foregrounds. In particular, \citet{Man2019} have shown that ignoring effects like LOS frequency decorrelation can bias $r$ determinations at consequential levels for next-generation experiments even if a frequency decorrelation parameter is used when fitting an ensemble of power spectra.

In this work, we focus on LOS frequency decorrelation, adopting a different approach, based on the fact that regions of the sky where the effect is expected to be important can be astrophysically identified using ancillary ISM data. 
Specifically, we use \hi emission data to identify sightlines that are potentially most susceptible to this effect. We combine information on the discrete number of \hi clouds on each sightline \citep{Pan2020} with an estimate of the magnetic field orientation in each cloud inferred from the morphology of linear \hi structures \citep{Cla2019}. This entirely \HI-based sample selection is agnostic to the {\it Planck} dust polarization data. We then compare the difference in polarization angles at 353 and 217~GHz along sightlines with and without an expected LOS frequency decorrelation effect, finding that the \hi data indeed identify sightlines with more significant EVPA rotation. This is the first detection of LOS frequency decorrelation with {\it Planck} data and illustrates the power of ancillary data such as \hi and stellar polarizations, to identify regions of the sky where the effect is most pronounced.

This paper is organized as follows. In Sect.~\ref{section:phenomenology} we briefly review the phenomenology of frequency decorrelation of polarization. In Sect.~\ref{sec:data} we describe the data sets that are used in the analysis. Section \ref{sec:analysis} presents the sample selection, the statistical tools that are used and our handling of biases and systematics. Section \ref{sec:results} presents our results. We discuss the robustness of our findings, and present further supporting observational evidence in Sect.~\ref{sec:validation}. An estimate of the required SED variation to reproduce the observed magnitude of LOS frequency decorrelation is presented in Sect.~\ref{sec:SEDVar}. We discuss our findings in Sect.~\ref{sec:discussion} and conclude in Sect.~\ref{sec:conclusions}.

This paper demonstrates that the effect of LOS frequency decorrelation exists at the pixel level and can be measured in the high-frequency polarization data from {\it Planck}. It does not address whether the amplitude of the effect is large enough at the sky-map level to affect any particular experiment's search for primordial B-modes. 

\section{Phenomenology of LOS frequency decorrelation}
\label{section:phenomenology}
We seek to detect LOS frequency decorrelation between \textit{Planck} polarization data at 353 and 217 GHz, frequencies dominated by Galactic thermal dust emission and the CMB.
Given that the polarized intensity of the CMB and of thermal dust emission feature different SEDs and that they are uncorrelated, their relative contribution to the observed polarization signal depends on the frequency. A change with frequency of the polarization position angle is therefore expected even if the polarization pattern of emission from dust remains constant across frequencies. Additionally, statistical and systematic errors induce scatter in polarization position angles at each frequency.
Therefore, a measured difference in polarization direction (electric vector position angle, EVPA) between frequencies cannot be immediately attributed to a LOS frequency decorrelation induced by multiple dust polarized-emission components.
Similarly, when the EVPA difference between frequencies is computed for a large statistical sample of different lines of sight, EVPA differences form a distribution with a finite spread. The three sources of EVPA differences mentioned above (noise, relative contributions of the CMB and the dust, and SED difference between dust components) each contribute to the width of the EVPA difference distribution.
We wish to detect a signal that {\em can} be directly attributed to frequency decorrelation of the dust polarized emission, in turn originating in the 3D structure of interstellar clouds and their magnetic field. Thus we have to construct a sample of lines of sight where dust decorrelation is expected to be significant, and then test whether the EVPA differences between frequencies are larger for that sample than for lines of sight where we expect that dust decorrelation is subdominant to effects from the CMB and noise.

The LOS frequency decorrelation of dust polarized emission is more likely to be observed for a given LOS if the following three conditions are met \citep{Tas2015}:
(i) at least two clouds are present along the LOS and both have a measurable emission contribution; (ii) the mean plane-of-sky magnetic field orientations of the clouds differ by an angle $\gtrsim 60^\circ$; (iii) the SEDs of the clouds are different. The first two conditions imply an emission with polarized intensity weaker than the sum of polarized intensities from individual clouds (LOS depolarization), and a modified polarization angle as compared to the emission from the dominant cloud. 
The third condition causes the polarization angle to be frequency dependent and is met if the dust clouds have different temperature and/or different polarization spectral index, e.g. if the dust grain properties differ between clouds.

In this work we rely on the fact that \hi column density correlates well with dust in the diffuse ISM (e.g. \citealt{Bou1996}; \citealt{PlaXI2014}; \citealt{Len2017}) and use recent \hi datasets to infer whether or not the aforementioned conditions are met.

\section{Data sets}
\label{sec:data}
In order to identify lines of sight where the LOS frequency decorrelation effect is most likely to be significant, we use two types of information that can be extracted from \hi observations. 
The first is the number of clouds along the LOS, obtained via a decomposition of \hi spectra by \cite{Pan2020}. We use publicly available\footnote{\url{https://doi.org/10.7910/DVN/8DA5LH}} results from this analysis to find sky pixels for which multiple clouds contribute to the dust emission signal in intensity.
The second is the plane-of-sky magnetic field orientation as a function of velocity, estimated via the morphology of \hi emission by \cite{Cla2019}. We use publicly available\footnote{\url{https://doi.org/10.7910/DVN/P41KDE}} results from this analysis to further constrain our pixel selection to lines of sight that contain clouds with significantly misaligned magnetic fields, i.e. the magnetic fields of the clouds form an angle with an absolute value between 60$^\circ$ and 90$^\circ$.
These \hi datasets allow us to define samples of sky pixels with which to study the sub-millimeter polarized emission as measured by {\it Planck}. We concentrate on the high-frequency {\it Planck} data, at 217 and 353 GHz, where thermal dust emission is known to dominate the measured polarization signal. 
In this section we describe the datasets that we use and the post-processing that we apply.

\begin{figure}
\begin{center}
\begin{tabular}{c}
$\mathcal{N}_c$ \\[-.2ex]
\includegraphics[trim={0.4cm 1.4cm 0.4cm .8cm},clip,width=.98\columnwidth]{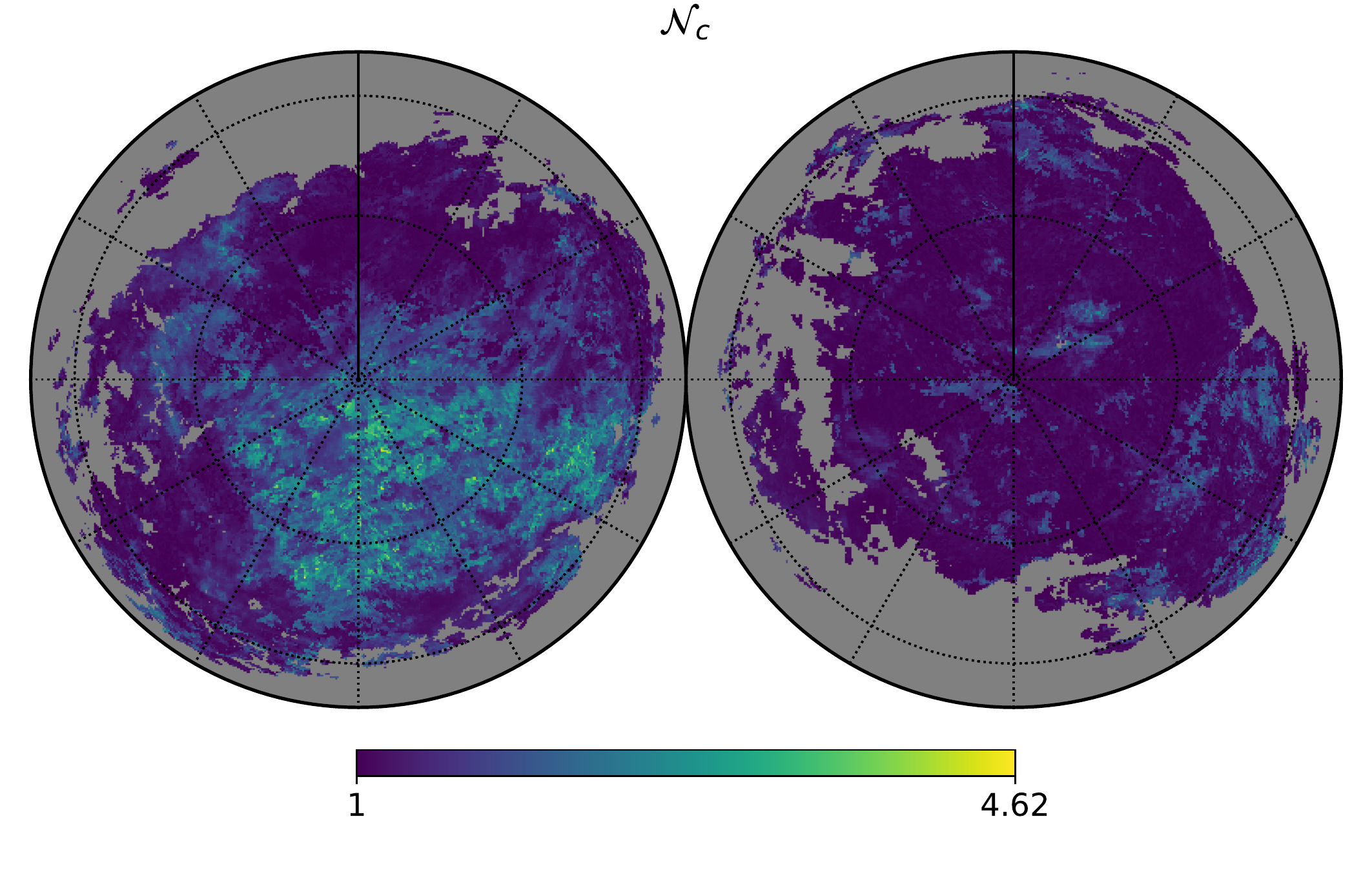}
\\[-.5ex]
{\small \hspace{.0cm} $1$ \hspace{3.8cm} $4.62$}
\\
\\[-.2ex]
$\Delta(\theta_{IVC},\theta_{LVC})$ \\[-.2ex]
\includegraphics[trim={0.4cm 1.4cm 0.4cm .8cm},clip,width=.98\columnwidth]{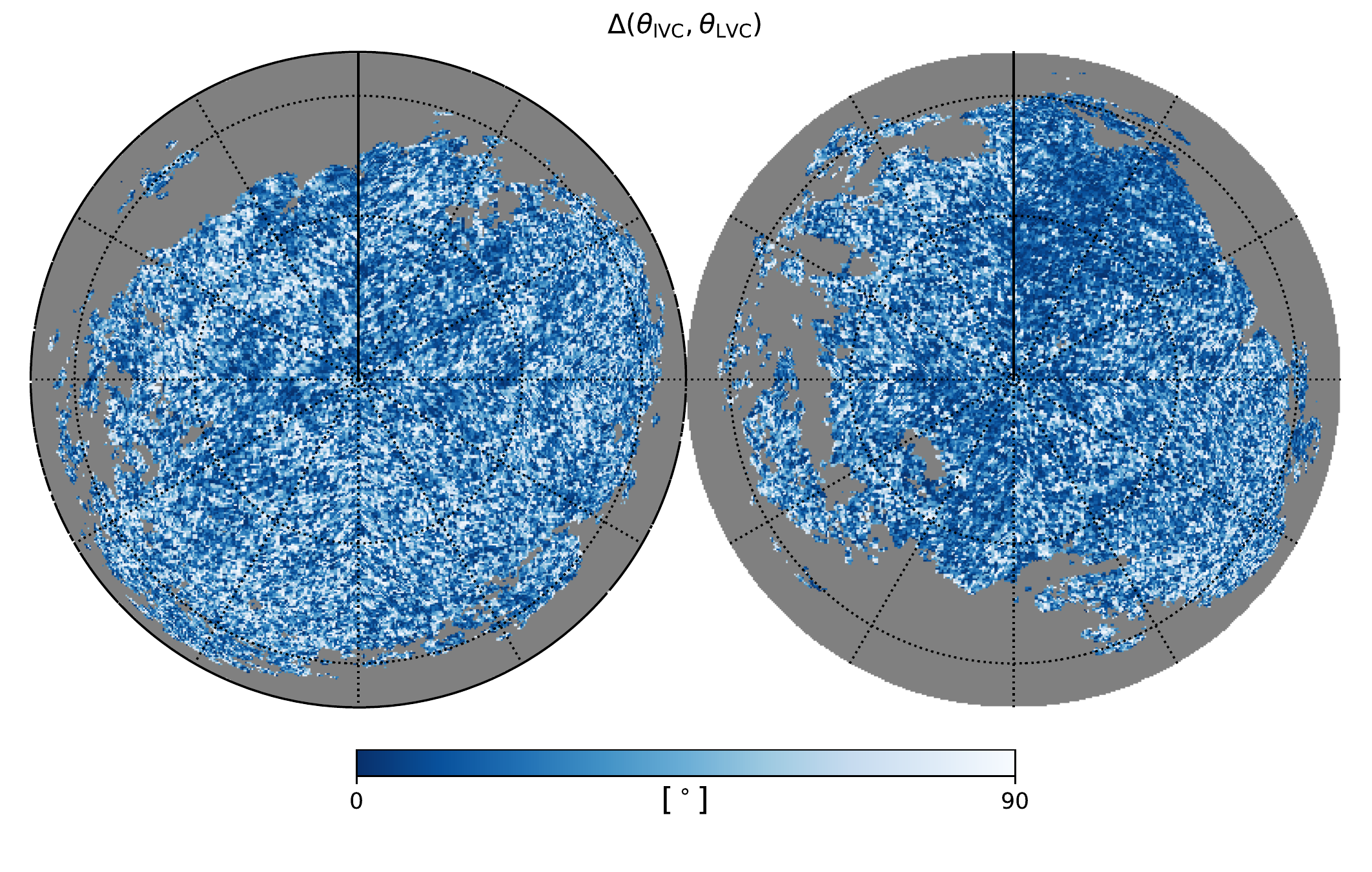}
\\[-.5ex]
{\small \hspace{.0cm} $0$ \hspace{1.75cm} [$^\circ$] \hspace{1.75cm} $90$}
\\
\\[-.2ex]
Sky positions \\[-.2ex]
\includegraphics[trim={0.4cm 1.48cm .4cm .8cm},clip,width=.98\columnwidth]{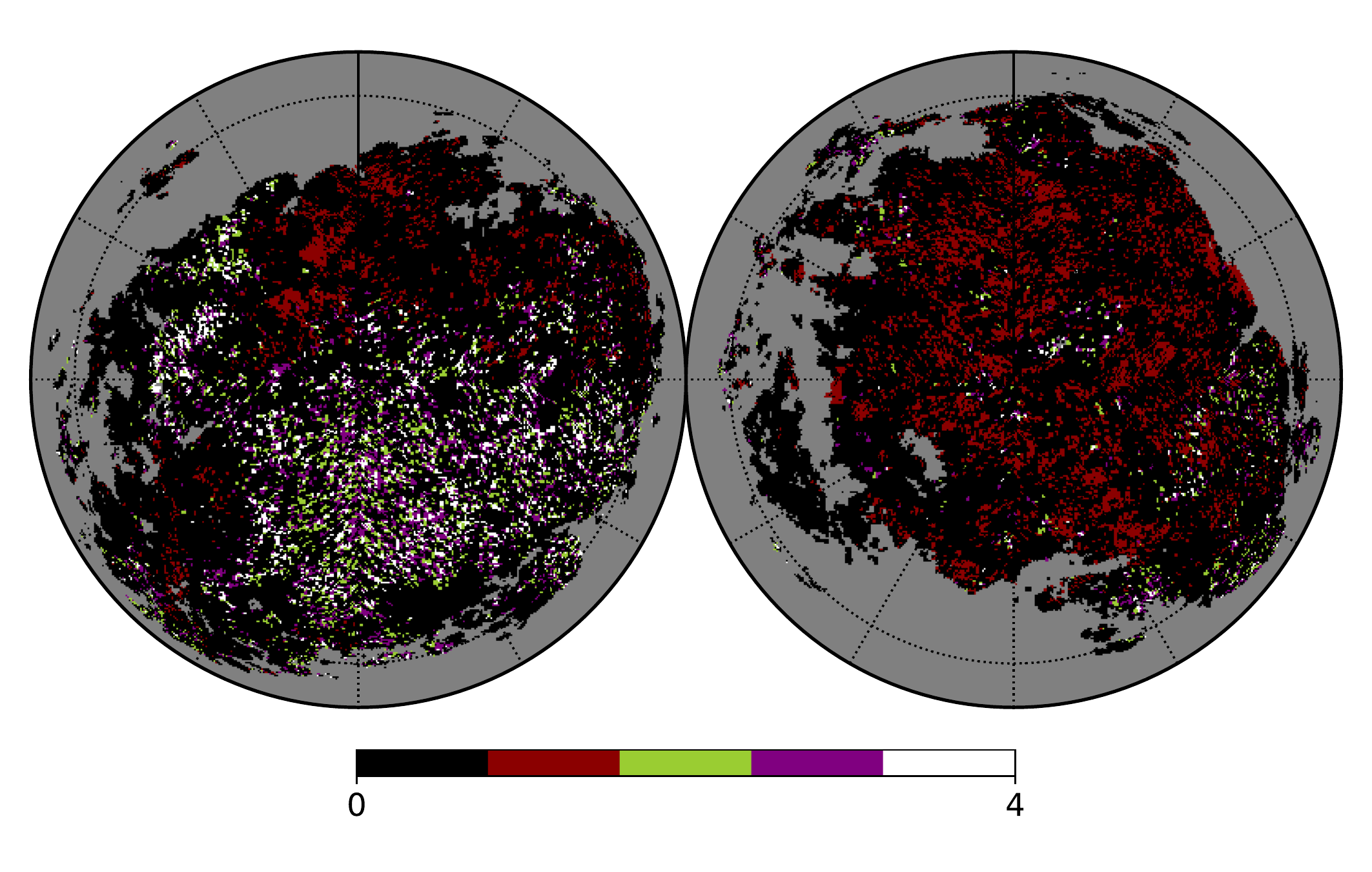}
\end{tabular}
\caption{Orthographic projections in Galactic coordinates. Longitude zero is marked by the vertical thick lines. The Galactic poles are at the centers of each disks. Galactic longitude increases counter-clockwise in the northern hemisphere (left) and clockwise in the southern one (right).
We show the maps of effective number of clouds $\mathcal{N}_c$ (top), and the map of $\Delta(\theta_{IVC},\theta_{LVC})$, used in  'Implementation 1' (second row).
Map of sky positions of pixel samples from 'Implementation 1' and 'Implementation 2' (bottom). White pixels are both in \texttt{target1} and \texttt{target2} samples. Green pixels are \texttt{target2} pixel not in \texttt{target1} and purple pixels are \texttt{target1} not in \texttt{target2}. Red pixels belong to \texttt{control} sample. Black pixels are those that belong to \texttt{all} but neither to \texttt{control} nor in \texttt{target1} or \texttt{target2}.}
\label{fig:dataMap-orth_Nc15}
\end{center}
\end{figure}

\subsection{\hi velocity components along the line of sight}
If multiple components of dust lie along the line of sight, and have different bulk kinematic properties, then the emission spectrum of the \hi line will show multiple peaks at different velocities with respect to the observer. This property of \hi emission was used by \cite{Pan2020} to measure the number of clouds along the LOS. The authors developed a method to identify the number of peaks in \hi spectra and applied it to data from the \HI4PI survey \citep{HI4PI2016} over the high Galactic latitude sky. The analysed area covers the parts of the sky where \hi column density is well correlated with far infrared dust emission, as defined by \cite{Len2017}. 

\cite{Pan2020} decomposed each \hi spectrum into a set of Gaussian components. The Gaussian parameters were grouped within HEALPix pixels of $N_{\rm{side}} = 128$ (termed `superpixels'), in order to construct a probability distribution function (PDF) of the components' centroid velocity. The PDFs were smoothed at a velocity resolution of 5 km s$^{-1}$. Within each superpixel, clouds were identified as kinematically distinct peaks in the PDF of Gaussian centroid velocity. The Gaussian components belonging to each peak were used to construct a velocity spectrum for each cloud. The published data products  include: (a) the column density of each cloud, $N_{{\rm HI}}$ and (b) the first and second moments of each cloud's spectrum ($v_0$, $\sigma_0$, respectively).

In sightlines with multiple components, not all components will contribute equally to the column density (and similarly to the total dust intensity). \cite{Pan2020} introduced a measure of the number of clouds per LOS that takes into account the column densities of clouds, defined as:
\begin{equation}
\mathcal{N}_c = \sum_i (N^i_{{\rm HI}})/N_{{\rm HI}}^{\rm max}
\end{equation}
where $N^i_{{\rm HI}}$ is the column density of the $i$-th cloud in the superpixel and $N_{{\rm HI}}^{\rm{max}}$ is the column density of the cloud with the highest $N_{{\rm HI}}$ in the superpixel. If the column density of a single cloud dominates the total column density of a superpixel, then $\mathcal{N}_c \sim 1$. If there are two clouds with equal column density, then $\mathcal{N}_c = 2$.

In this paper we use $\mathcal{N}_c$, a map of which is shown in Fig.~\ref{fig:dataMap-orth_Nc15} (top), to distinguish between sightlines whose dust emission is dominated by a single component and those where multiple components might be contributing to the signal.
\cite{Pan2020} have shown that $\mathcal{N}_c$ is anticorrelated with the degree of linear polarization at 353~GHz, suggesting that lines of sight where multiple components contribute to the polarization signal exhibit larger LOS depolarization than the rest of the sky.
However, a simple selection on $\mathcal{N}_c$ alone does not imply a high ratio of column densities between clouds; a value of $\mathcal{N}_c = 1.5$ can be achieved by two clouds or by an arbitrary number of clouds, the former case being in general more likely to induce measurable LOS frequency decorrelation.
Thus in one variation of our pixel selection we consider a different metric 
(see Sect.~\ref{sec:analysis}) involving the ratio of dominant cloud column densities, $\mathcal{F}_{21}$, defined as follows: 
for pixels with at least two clouds ($\mathcal{N}_c > 1$), 
\begin{equation}
\label{eq:F}
    \mathcal{F}_{21} = N_{{\rm HI}}^{\rm{max2}}/N_{{\rm HI}}^{\rm{max}},
\end{equation}
where $N_{{\rm HI}}^{\rm{max}}$ is the column density of the cloud with the highest $N_{{\rm HI}}$, and $N_{{\rm HI}}^{\rm{max2}}$ is that of the cloud with second-highest $N_{{\rm HI}}$.
We use the cloud column densities provided by \citet{Pan2020}.

\medskip

\subsection{Orientation of \hi structures}
The morphology of \hi emission encodes properties of the ambient magnetic field in two measurable ways. First, high-resolution \hi channel maps reveal thin, linear structures that are well aligned with the magnetic field as traced by starlight polarization \citep{Cla2014} and polarized dust emission \citep{Cla2015, Mar2015}.
These magnetically aligned \hi structures are associated with anisotropic cold \hi gas (\citealt{McC2006}; \citealt{Cla2019a}; \citealt{Pee2019}; \citealt{Kal2020}; \citealt{Mur2020}). Second, the degree of alignment of linear \hi structures as a function of LOS velocity traces LOS magnetic field tangling, and therefore the observed dust polarization fraction \citep{Cla2018}.

These insights were synthesized into a formalism by \cite{Cla2019} that defines 3D maps of the Stokes parameters of linear polarization. These maps are based purely on the morphology of \hi emission. The distribution of linear \hi emission as a function of orientation on the sky is quantified by the Rolling Hough Transform \citep[RHT;][]{Cla2014}. The RHT is applied to discrete \hi velocity channels in an \hi data cube to calculate maps of $R(v, \theta)$, the linear intensity as a function of line-of-sight velocity $v$ and orientation $\theta$. $R(v, \theta)$ is normalized such that it can be treated analogously to a probability distribution function for the orientation of \hi in each pixel. The \HI-based Stokes parameters are then defined as: 

\begin{equation}
    Q_\mathrm{HI}(v) = I_\mathrm{HI}(v)\sum_\theta R(v, \theta) \cos(2\theta) d\theta
\label{eq:Q_HI}
\end{equation}
\begin{equation}
    U_\mathrm{HI}(v) = I_\mathrm{HI}(v)\sum_\theta R(v, \theta) \sin(2\theta) d\theta,
\label{eq:U_HI}
\end{equation}
where $I_\mathrm{HI}(v)$ is the \hi intensity as a function of LOS velocity.
Integrating $Q_\mathrm{HI}(v)$ and $U_\mathrm{HI}(v)$ over the velocity dimension yields \HI-based Stokes $Q_\mathrm{HI}$ and $U_\mathrm{HI}$ maps that reproduce the \textit{Planck} 353 GHz $Q$ and $U$ maps with remarkable fidelity. \citet{Cla2019} also demonstrate consistency with a tomographic determination of the magnetic field orientation along one line of sight based on measurements of optical starlight polarization and \textit{Gaia} stellar distances \citep{Pan2019}.

We therefore use the \citet{Cla2019} maps as a probe of the local magnetic field orientation as a function of LOS velocity. We use their \HI4PI-based maps, which use a non-uniform LOS velocity bin size and cover the full sky at the \HI4PI angular resolution of $16.2'$ (see \citet{Cla2019} for map details). To match the resolution and pixelization of the $N_c$ map, we apply a Gaussian filter to degrade the Clark \& Hensley maps to a uniform 30\arcmin 
~resolution, and use the \texttt{healpy} function \texttt{ud\_grade} to bin the smoothed maps to $N_{\rm{side}} = 128$. We can use these 3D maps to measure the \HI-based polarization angle in a specified velocity range by summing $Q_\mathrm{HI}(v)$ and $U_\mathrm{HI}(v)$ over the desired velocity bins and computing $\theta_\mathrm{HI} = 1/2\, \arctan (-U_\mathrm{HI},\,Q_\mathrm{HI})$, where $\arctan$ is the 4-quadrant inverse tangent function here and throughout this paper.
In this paper we use $\theta$ to denote the position angle of \hi structures and $\psi$ for polarization position angles.

\subsection{Polarization data from the {\it Planck} satellite}\label{data}
In this work we employ two full-sky sets of sub-millimeter polarization data, both obtained by the {\it Planck} satellite.
First we utilize the third data release of the {\it Planck} collaboration (PR3). We use the 217 GHz single-frequency maps and the 353 GHz single-frequency maps from the polarization-sensitive bolometers only, as recommended in \cite{PlaIII2018} and \cite{PlaXII2018}, which we downloaded from the Planck Legacy Archive\footnote{\url{http://pla.esac.esa.int}} (PLA).

Second, we use a more recent set of high-frequency polarization maps obtained from {\it Planck} data but processed through the upgraded map-making algorithm \texttt{SRoll2} that corrects data for known residual systematics in Legacy maps down to the detector noise level (\citealt{Del2019}).
We use the full-dataset Polarization Sensitive Bolometers \texttt{SRoll2} polarization maps at frequency 353 and 217 GHz available at their website\footnote{\url{http://sroll20.ias.u-psud.fr/sroll20_data.html}}.
We note that most of the analysis presented in this paper was completed before the \texttt{Npipe} maps became available (\citealt{PlaLVII2020}). Analyzing this new set of maps would require the implementation of a different analysis pipeline than the one developed and used in this work because per-pixel block-diagonal covariance matrices are not available. However, we note that preliminary studies using \texttt{Npipe} maps yield results consistent with those obtained in this paper, in the direction of the detection of LOS frequency decorrelation being more significant than that obtained using PR3 maps.

We apply the same post-processing to both sets of polarization maps. We smooth the $I$, $Q$, and $U$ maps to a resolution of 30\arcmin ~in order to increase the signal-to-noise ratio.
We smooth the per-pixel block-diagonal polarization covariance matrices following the analytical prescription in Appendix A of \cite{PlaXIX2015}. This formalism neglects correlations between neighboring pixels, but takes into account the off-diagonal covariance between the $Q$ and $U$ Stokes parameters. These terms can be substantial at high Galactic latitudes. 

\smallskip

When necessary, we propagate the observational uncertainties in our analysis by making use of Monte Carlo (MC) realizations of correlated noise using a Cholesky decomposition of the smoothed per-pixel block-diagonal covariance matrix (see e.g. Appendix A of \cite{PlaXIX2015} or Appendix B of \cite{Ska2019}). To assess the observational uncertainty on a measurement, we repeat our analysis on those simulated Stokes parameters and study the resulting per-pixel distribution. We validated this approach by comparing to analytical estimates the uncertainties obtained for the polarized intensity and the polarization position angle.

\begin{figure*}
    \centering
    \includegraphics[width=\textwidth]{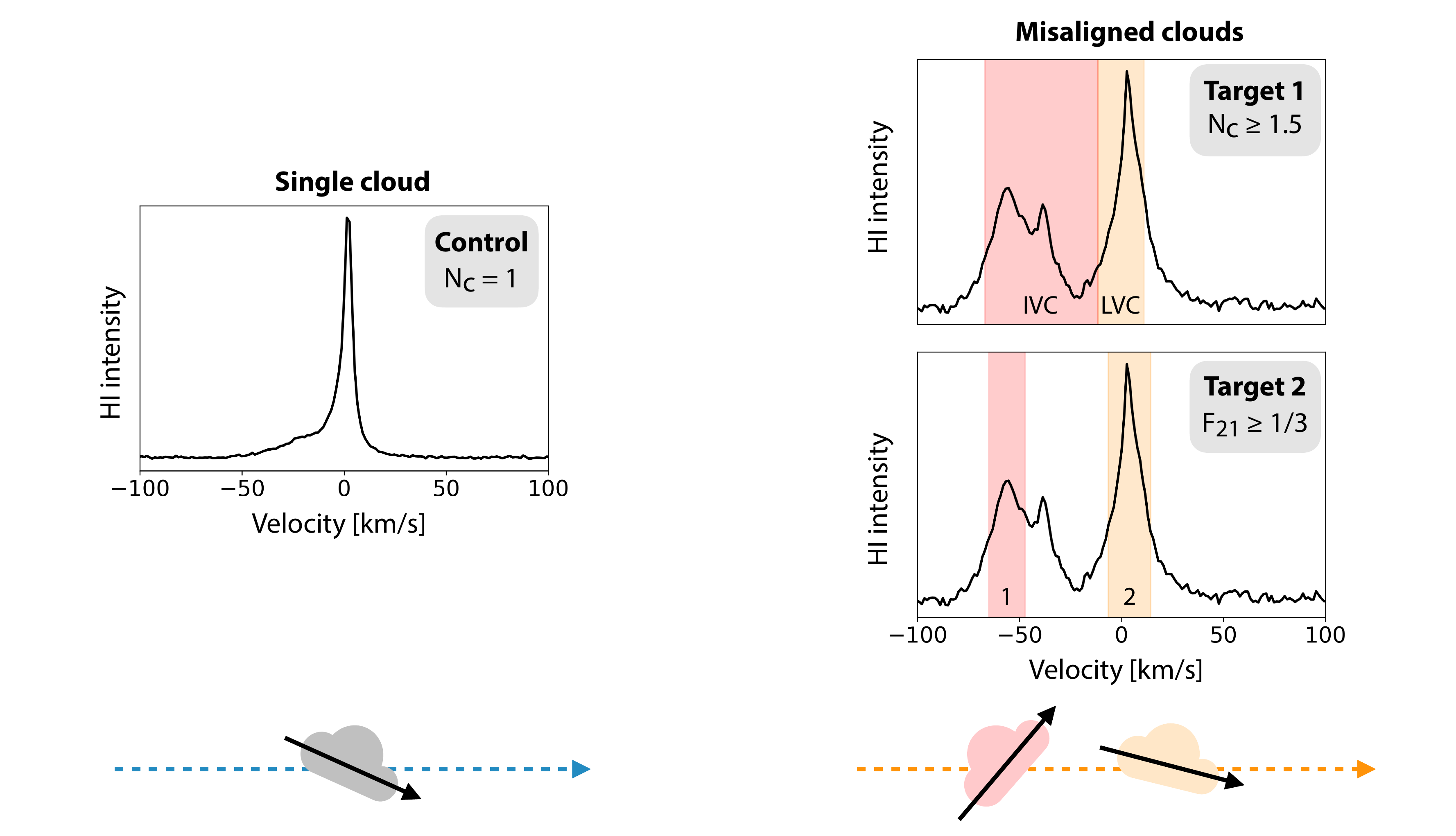}
    \caption{Cartoon illustration of the pixel selection described in Sect.~\ref{subsec:pixelsamples}. Left panel: \hi intensity spectrum of a representative pixel from our \texttt{control} group. The \texttt{control} sample targets sightlines defined by a single \hi cloud, parameterized by $\mathcal{N}_c = 1$. Right panel: \hi intensity spectrum of a representative pixel that is included in both \texttt{target1} and \texttt{target2}. Pixels in the \texttt{target} samples are selected to have multiple \hi clouds along the line of sight, as parameterized by either $\mathcal{N}_c \geq 1.5$ (\texttt{target1}) or $\mathcal{F}_{21} \geq 1/3$ (\texttt{target2}). \hi orientations are determined for two clouds along each target line of sight by summing the \cite{Cla2019} \HI-based Stokes parameters over the indicated velocity ranges, and we require that the angles in these clouds differ by at least 60$^\circ$. Cloud orientations in the \texttt{target1} sample are determined from predefined IVC and LVC velocity ranges. Cloud orientations in the \texttt{target2} sample are determined from the $1\sigma$ velocity range around the two most prominent \hi clouds identified in \cite{Pan2020}. }
    \label{fig:cartoon}
\end{figure*}

\subsection{CMB polarization maps}
We make use of the CMB polarization maps obtained from the four component-separation algorithms used by the Planck Collaboration, and applied to the third release of the {\it Planck} data: \texttt{commander}, \texttt{nilc}, \texttt{sevem}, and \texttt{smica} (\citealt{PlaXII2014}, \citealt{PlaIX2016}, \citealt{PlaIV2018}, and references therein).
We downloaded the CMB maps from the PLA and smoothed them so that they all have an effective resolution corresponding to a Gaussian beam with FWHM of 30\arcmin, just as we do with the single-frequency maps used in this work.

\section{Analysis Framework}
\label{sec:analysis}

\subsection{Sample selection}
\label{subsec:pixelsamples}

In order to determine statistically if LOS frequency decorrelation is present and measurable in the {\it Planck} high-frequency polarization data, we construct astrophysically-selected samples of pixels on the sky based only on \hi data.

We distinguish between our samples using the labels \texttt{all}  \citep[all the pixels in the high Galactic latitude LOS cloud decomposition of][]{Pan2020}; \texttt{control} (pixels that should not exhibit LOS frequency decorrelation); and \texttt{target} (pixels that are likely to exhibit large LOS frequency decorrelation).
According to \cite{Tas2015}, the degree of LOS decorrelation between two frequencies depends on (a) how the ratio of polarized intensities contributed by distinct components along a LOS changes between frequencies; and (b) the degree of magnetic field misalignment between these contributing components. The first factor above depends non-trivially on both the temperature difference between components, and on the amount of emitting dust (column density) in each. Our physical understanding of these dependencies motivates our definition of \texttt{control} and \texttt{target} samples from \hi data:
\begin{itemize}
    \item[-]\texttt{control}: If the dust emission is strongly dominated by a single component (cloud), no LOS frequency decorrelation is expected, regardless of the other criteria above. For this reason, we construct our \texttt{control} sample using \hi data to select pixels where a single component dominates the \hi emission (proxy for the emitting dust).
    \item[-]\texttt{target}: For LOS frequency decorrelation to be significant, there must be (a) more than one contributing component, and (b) a significant misalignment ($\gtrsim 60^\circ$) between the orientations of plane-of-sky magnetic field that permeate the components. Both criteria are required for a pixel to be included in the \texttt{target} sample.
    We do not attempt to use the \hi data to make predictions about the shape of the dust SED.
\end{itemize}
The nature of the \texttt{control} sample allows for a simple selection criterion: requiring that pixels contain a single cloud along the LOS. We therefore select those pixels that have a column-density-weighted number of clouds (see Sect.~\ref{sec:data}) equal to unity ($\mathcal{N}_c = 1$). For the \texttt{target} sample, however, there exist different ways in which these selection criteria can be implemented in practice. For this reason, we have performed the analysis using two distinct implementations of the sample selection, so as to ensure that our particular choices do not qualitatively affect our results. Our selection criteria are described below and are summarized in Table~\ref{tab:sampleDef} and Fig.~\ref{fig:cartoon}.

\begin{table}
\caption{Criteria to define the samples in Implementation 1 and Implementation 2.}
\label{tab:sampleDef}
\centering
\begin{tabular}{l l l}
\hline\hline
\\[-1.5ex]
    &   {Implementation 1}  &  {Implementation 2} \\
\\[-1.5ex]
\hline \\[-1.5ex]
\texttt{control}    & $\mathcal{N}_c = 1$   & $\mathcal{N}_c = 1$ \\
\\[-1.5ex]
\texttt{target}
     & $\mathcal{N}_c \geq 1.5$ & $\mathcal{F}_{21} \geq 1/3$ \\
   & $\Delta(\theta_{IVC},\theta_{LVC}) \geq 60^\circ$
    & $\Delta(\theta_{1},\theta_{2}) \geq 60^\circ$ \\
\\[-1.5ex]
\hline
\end{tabular}
\end{table}
 
{\bf Implementation 1:}
The first criterion for constructing the \texttt{target} sample in this implementation (hereafter \texttt{target1}) selects pixels for which $\mathcal{N}_c \geq 1.5$.

This ensures that there is a significant contribution to the dust emission signal in intensity that is not from the dominant component.
The same will hold for polarized intensity, with the exception of special cases where the magnetic field in one of the clouds lies mainly along the line of sight (which would result in very little, if any, polarized emission from the specific cloud). While we cannot control for the unknown 3D geometry of the magnetic field in each cloud, this unknown simply adds noise to the LOS frequency decorrelation signal we are after -- our selection of a statistically large sample of pixels likely contains all possible relative orientations between the 3D magnetic field of clouds along the same LOS.

In addition to the requirement that $\mathcal{N}_c \geq 1.5$, \texttt{target1} pixels must also satisfy a misalignment condition. To impose such a condition, we first post-process the \citet{Cla2019} \HI-based Stokes parameter data (provided in pre-defined discrete velocity bins) to obtain orientation information on a per-cloud basis. 
We make use of the commonly used distinction of high-latitude \hi clouds with respect to their velocity: Low Velocity Clouds are found in the range $-12 \, {\rm{km\,s^{-1}}} \leq v_0 \leq 10\, {\rm{km\,s^{-1}}}$ while Intermediate Velocity Clouds (IVC) are found in the range $-70\, {\rm{km\,s^{-1}}} \leq v_0 \leq -12\, {\rm{km\,s^{-1}}}$ or $10\, {\rm{km\,s^{-1}}} \leq v_0 \leq 70\, {\rm{km\,s^{-1}}}$ (where $v_0$ is the cloud centroid velocity and the velocity ranges are defined as in \citealt{Pan2020}). These two classes of clouds are found to show systematic differences in their dust properties, with IVCs, for example, having higher dust temperatures than LVCs on average (e.g., \citealt{PlaXXIV2011}; \citealt{PlaXI2014}; \citealt{Pan2020}). Pixels in which the \hi orientation changes significantly between the LVC and IVC range likely satisfy all necessary conditions for the LOS frequency decorrelation effect: varying dust SED \textit{and} magnetic field orientation along the LOS (in addition to the requirement of $\mathcal{N}_c \geq 1.5$). 

For each pixel we thus compute the orientation of two `effective' clouds: an LVC and an IVC. For this we sum the \hi Stokes parameters within the LVC and IVC velocity ranges separately, and then calculate a single \hi orientation within the LVC range, $\theta_{LVC}$, and within the IVC range, $\theta_{IVC}$. 
For a pixel to be included in the \texttt{target1} sample, the misalignment criterion requires that the angles $\theta_{LVC}$ and $\theta_{IVC}$ differ by at least 60$^\circ$.
The (unsigned) angle difference between two angles expressed in radians is computed as
\begin{equation}
    \Delta(\xi_1,\, \xi_2) = \pi/2 - |\pi/2 - |\xi_1 - \xi_2||
\label{eq:Delta}
\end{equation}
where $\xi_{1,2}$ are position angles (either $\theta$'s or $\psi$'s) defined in the range $\left[0,\pi\right)$ and
where the consecutive absolute values take into account the $\pi$ degeneracy of orientations.

\smallskip

{\bf Implementation 2:} We modify the criteria for constructing the \texttt{target} sample in order to test for the robustness of our results against sample selection. First, we identify pixels with {\em at least two} significant \hi components by requiring (a) that $\mathcal{N}_c > 1$ and (b)  the ratio of column densities of the two main \hi components, $\mathcal{F}_{21}$, is high (see Eq.~\ref{eq:F}). Specifically, 
candidate pixels for the \texttt{target} sample in this implementation (hereafter \texttt{target2}) are selected so the column density of the second most prominent component is at least one third of the dominant component, i.e.  $\mathcal{F}_{21} \geq 1/3$. By using $\mathcal{F}_{21}$ instead of a higher threshold in the value of $\mathcal{N}_c$ (as was done in Implementation 1) we ensure that the dust emission signal (in intensity at least) arises mainly from 2 clouds of comparable $N_{\rm{HI}}$, rather than a larger number of low-$N_{\rm{HI}}$ clouds (as discussed in Sect.~\ref{sec:data}).

We also modify the construction of the per-cloud \hi orientation, compared to Implementation 1. For each cloud, we consider the velocity range within $v_0 \pm \sigma_0$, where $v_0$ is the cloud centroid velocity and $\sigma_0$ is the second moment of its spectrum. We sum the \hi Stokes parameters of the Clark \& Hensley maps within this velocity range creating maps of per-cloud Stokes parameters, $Q_{\rm{HI}}^\mathrm{cloud}$ and $U_{\rm{HI}}^\mathrm{cloud}$. For each pixel we use these per-cloud Stokes parameters to calculate the \hi orientation of the highest-$N_{\rm{HI}}$ cloud, $\theta_1$, and that of the second highest-$N_{\rm{HI}}$ cloud, $\theta_2$. The \texttt{target2} sample is constructed by requiring pixels to have $\Delta(\theta_{1},\,\theta_{2}) \geq 60^\circ$, in addition to the aforementioned column-density-based criteria. This cloud-based definition of the misalignment condition avoids relying on the predefined velocity ranges for the LVC and IVC components.

\medskip

\paragraph{\bf Statistical properties of the samples:}
The samples contain $N_{\rm{all}} = 83374$, $N_{\rm{control}} = 7328$, $N_{\rm{target1}} = 5059$, and $N_{\rm{target2}} = 5755$ high-latitude pixels on a HEALPix map (\citealt{Gor2005}) of $N_{\rm{side}} = 128$.
The pixels in \texttt{target1} (\texttt{target2}) represent about 6.1\% (6.9\%) of the high-latitude sky defined by the $\mathcal{N}_c$ data and about 2.6\% (2.9\%) of the full sky.
\texttt{target1} and \texttt{target2} have 2383 pixels in common. This overlap is to be expected, since despite the different specific criteria, both Implementations 1 and 2 are motivated by the same astrophysical requirements.

In Fig.~\ref{fig:dataMap-orth_Nc15} we show polar projections of the $\mathcal{N}_c$ map (top), the difference of position angle between the IVC and LVC effective clouds (second row), followed by sky position of the pixels of our \texttt{control} and \texttt{target1} and \texttt{target2} samples (bottom). We note that there is a significant difference between the locations of \texttt{target} and \texttt{control} pixels: the former are preferentially found in the northern hemisphere (in both implementations), while the latter are mostly found in the southern hemisphere. This uneven distribution is inherited from the spatial distribution of $\mathcal{N}_c$. As noted in \cite{Pan2020}, $\mathcal{N}_c$ is spatially correlated with the column density of IVCs. The presence of these clouds primarily in the northern hemisphere has been noted already from earlier studies of Galactic \hi surveys (e.g. \citealt{Danly89}; \citealt{KuntzDanly96}), and is tied to their astrophysical origin (e.g. \citealt{ShapiroField76}; \citealt{Bregman80}; \citealt{WessFejes73}; \citealt{Heiles84}; \citealt{Verschuur93}).

\subsection{Statistical Methodology}
\label{subsec:stats}
We select pixels from the {\it Planck} 353 and 217 GHz polarization maps for each of our three samples, and compute the signed-difference between the EVPAs according to
\begin{equation}
\begin{aligned}
    \Delta_s(\psi_{353},\psi_{217}) = \frac{1}{2}\, \arctan ({\sin \left[ 2\, (\psi_{353} - \psi_{217}) \right]} , \, \\
{\cos \left[ 2\, ( \psi_{353} - \psi_{217}) \right]})
\end{aligned}
\label{eq:Delta_s}
\end{equation}
where the EVPA at both frequencies is determined from the Stokes $Q_\nu$ and $U_\nu$ according to $\psi_\nu = 1/2\, \arctan (-U_\nu,\,Q_\nu)$ and has a value in the range $[0^\circ,\, 180^\circ$).
$\Delta_s(\psi_{353},\psi_{217})$ is defined in the range $[-90^\circ,\, 90^\circ]$.
The subscript $s$ in $\Delta_s$ is used to denote the signed difference of EVPA from Eq.~\ref{eq:Delta}, the un-signed position angle difference (the two are related through $\Delta(\xi_1,\xi_2) = |\Delta_s(\xi_1,\xi_2)|$).

We choose to use the signed angle difference rather than the unsigned version because an ensemble of signed angle differences is centered on and symmetric about zero in the absence of systematic offsets.
For an ensemble of $N$ 2-circular quantities $\{\xi_{1,2,...,N}\}$, the circular mean and the circular standard deviation are defined as
\begin{equation}
\left\langle \{\xi\} \right\rangle = \frac{1}{2}\,\arctan \left( {\sum_{n=1}^N \sin (2\xi_n)},\, {\sum_{n=1}^N \cos (2\xi_n)} \right)
\label{eq:CircMean}
\end{equation}
and
\begin{equation}
S(\{\xi\}) = \sqrt{ - \log \left[ \left( \frac{1}{N} \sum_{n=1}^{N} \sin (2\xi_n) \right)^2 +  \left(\frac{1}{N} \sum_{n=1}^{N} \cos (2\xi_n) \right)^2 \right] } \; .
\label{eq:CircStd}
\end{equation}

For a sample of pixels the distribution of $\Delta_s(\psi_{353},\psi_{217})$ is expected to have a circular mean close to zero and a finite circular standard deviation. The latter encodes a decorrelation of EVPAs between frequencies due to ({\it i}) uncorrelated noise at different frequencies; ({\it ii})
the relative contribution of dust and CMB at the two frequencies; 
and ({\it iii}) LOS frequency decorrelation due to the polarized intensity contribution from distinct misaligned dust clouds with SEDs varying between frequencies (the effect we are seeking to detect).

Because the \texttt{target} samples are selected to have a higher likelihood of large LOS frequency decorrelation, we predict a larger circular standard deviation for the \texttt{target} sample than for the \texttt{control} sample. 
Therefore we adopt the spread of the distribution of polarization angle differences as our test statistic:
\begin{equation}
\label{eq:D}
\mathcal{D} \equiv S(\{\Delta_s(\psi_{353},\psi_{217})\}),
\end{equation}
where a detection of LOS frequency decorrelation would correspond to a larger $\mathcal{D}$ for the \texttt{target} sample than for \texttt{control}.
Any inference of the presence of LOS frequency decorrelation has to account for the other sources of increased scatter in the distribution of $\Delta_s(\psi_{353},\psi_{217})$, i.e. residual systematics, CMB polarization, sampling uncertainties, and data noise, and must consider the possibility that these properties differ between \texttt{target} and \texttt{control}.

\smallskip

We address the first two effects (residual systematics and CMB polarization) by repeating our analysis on maps that are derived from the same raw {\it Planck} data, but processed differently.
\begin{figure}
\begin{center}
\begin{tabular}{c}
\hspace{1.1cm} Difference between PR3 and \texttt{SRoll2} maps \\[-.2ex]
\rotatebox{90}{\hspace{1.8cm} Normalized Distribution}
\includegraphics[trim={1.cm .6cm .2cm .3cm},clip,width=.95\columnwidth]{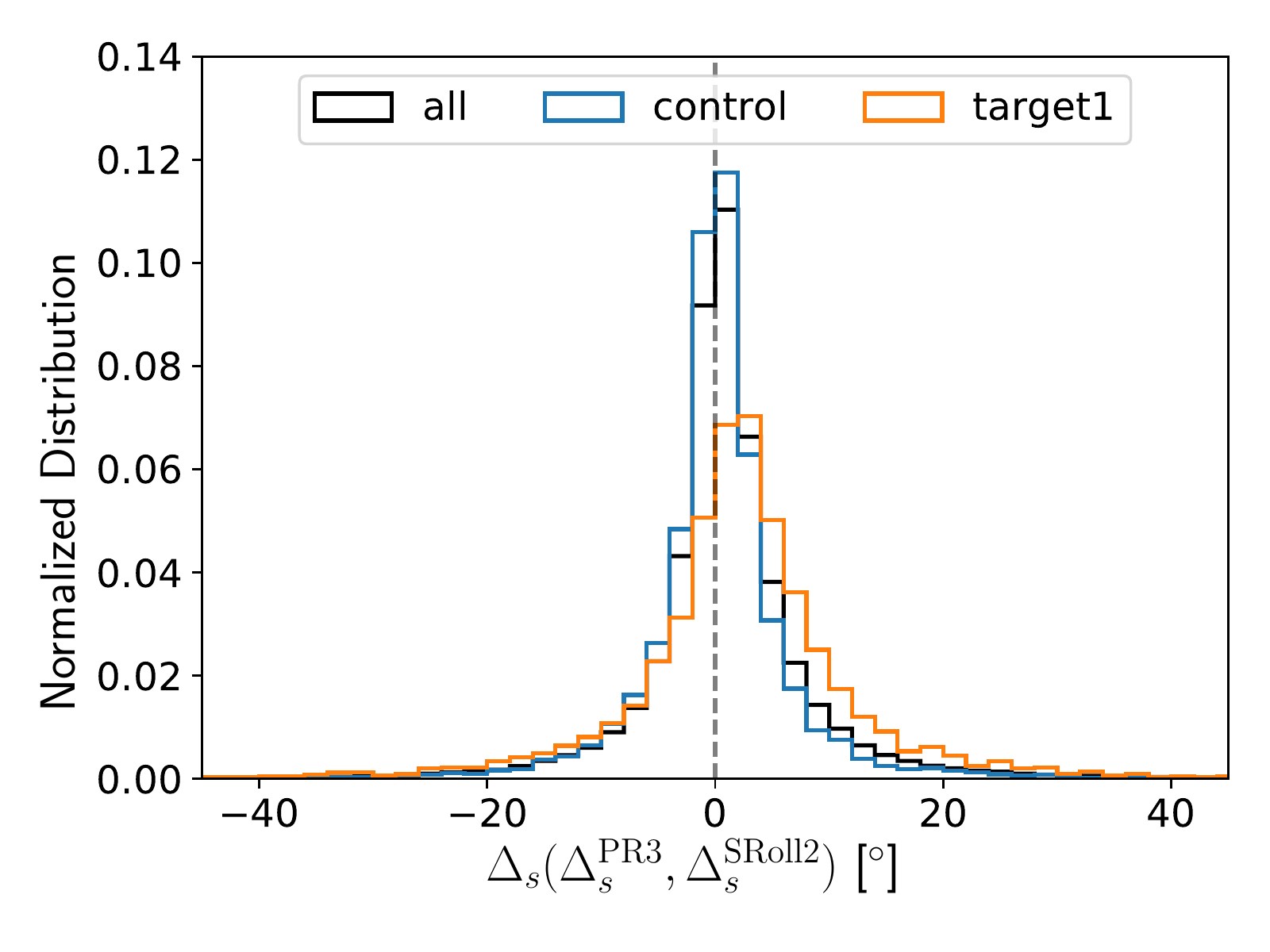}
\end{tabular}
\caption{Normalized histogram of the difference between PR3 and \texttt{SRoll2} maps of the EVPA difference between 353 and 217 GHz [$\Delta_s(\Delta_s(\psi_{353},\psi_{217})^{\rm{PR3}},\Delta_s(\psi_{353},\psi_{217})^{\rm{SRoll2}})$] for sky pixels of \texttt{all}, \texttt{control} and \texttt{target1}. For most pixels, the results agree within $\sim \pm 5^\circ$; however pixels of \texttt{target1} exhibit larger differences between map versions, centered at $2.3^\circ$. This suggests that the sky area covered by our \texttt{target1} sample received more correction from the systematic cleaning. A similar picture is obtained considering \texttt{target2} instead of \texttt{target1}. 
}
\label{fig:PR3vsSRoll2_DPPAs}
\end{center}
\end{figure}
One plausible concern is that spatially correlated systematics in PR3 maps affect \texttt{target} and \texttt{control} differently, resulting in a false-positive detection of LOS frequency decorrelation. To exclude this possibility, we repeat our analysis using the improved version of {\it Planck} HFI polarization maps obtained from the \texttt{SRoll2} map-making algorithm that better corrects for known residual systematics down to the detector noise level (\citealt{Del2019}). The difference between the $\Delta_s(\psi_{353},\psi_{217})$ distributions computed from the PR3 and \texttt{SRoll2} maps particularized to our samples is shown in Fig.~\ref{fig:PR3vsSRoll2_DPPAs}. We find that this difference distribution is offset from 0 for the \texttt{target} samples, indicating that the region of sky containing the \texttt{target} pixels differed systematically between the PR3 and \texttt{SRoll2} maps; a conclusion also reached from inspection of Fig.~7 of \cite{Del2019}.

\begin{figure}
\begin{center}
\begin{tabular}{c}
\hspace{.6cm} Polarized intensity of pixels in each sample/frequency\\ \\[-1.3ex]
\multirow{2}{*}{\rotatebox{90}{\hspace{.4cm} Normalized Distribution\hspace{-3.8cm}}}
\includegraphics[trim={1.cm 1.2cm .1cm .4cm},clip,width=.96\columnwidth]{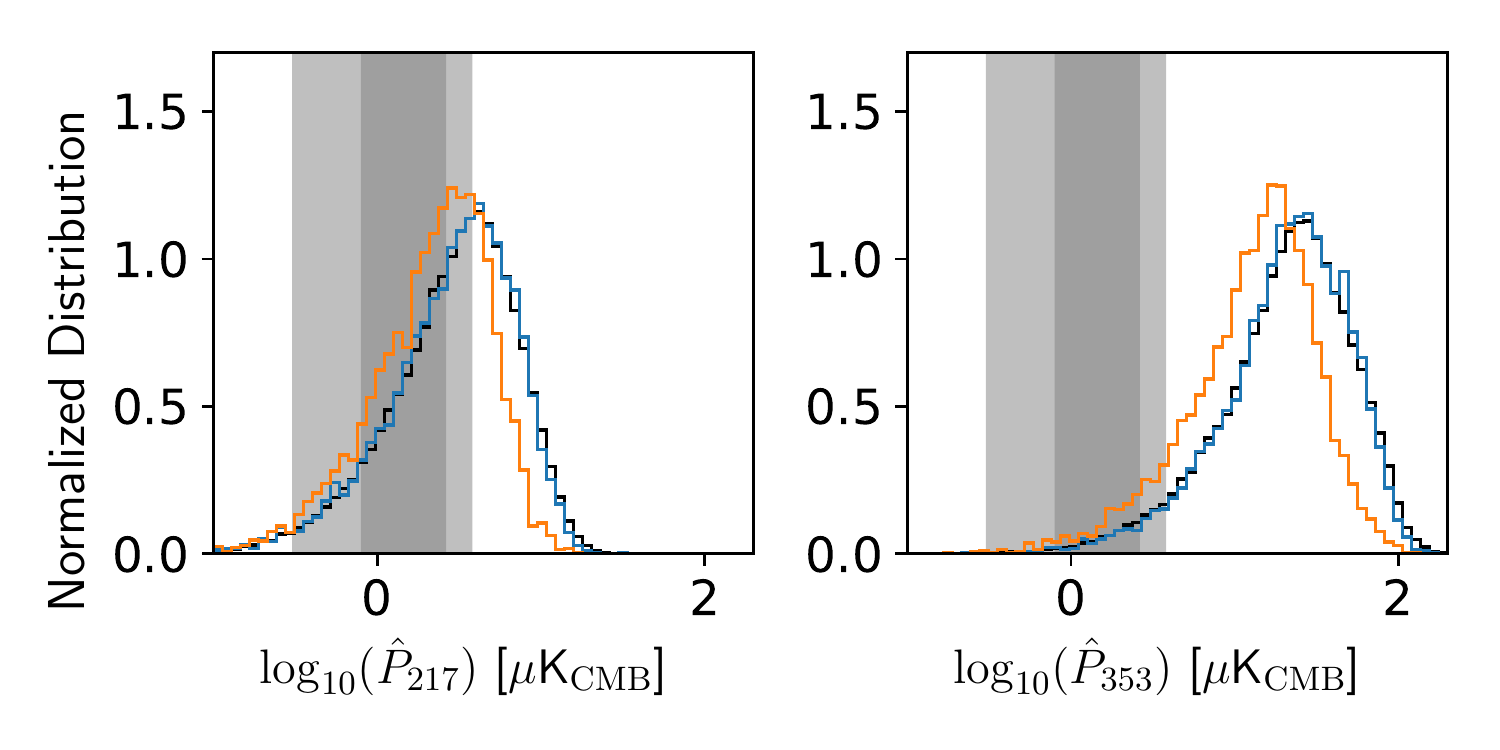}\\[-.5em]
{\small \hspace{.8cm}  $\log_{10}(\hat{P}_{217})$ [$\mu$K$_{\rm{CMB}}$]   \hspace{1.6cm} $\log_{10}(\hat{P}_{353})$ [$\mu$K$_{\rm{CMB}}$]} \hspace{.2cm}
\end{tabular}
\caption{
Histograms of debiased polarized intensity $\hat{P}$ at 217 GHz (left) and 353 GHz (right) (\citealt{Pla2014}) of \texttt{all} (black), \texttt{control} (blue) and \texttt{target1} (orange).
(Dark) gray shaded areas mark (68) 95 percent of the CMB contribution to the polarized intensity as inferred by \texttt{smica} for a FWHM beam of 30\arcmin ~and for the $\mathcal{N}_c$ footprint. The CMB contribution is negligible at 353 GHz but not at 217 GHz, especially for pixels of low $\hat{P}_{217}$.
Histograms correspond to PR3 polarization maps with no CMB subtraction. }
\label{fig:cmb_cont}
\end{center}
\end{figure}

\smallskip

A second plausible concern is that the contribution of the CMB to the polarized intensity changes between 353 GHz (where it is largely negligible) and 217 GHz (where it might be considerable, especially for pixels with low 217 GHz polarized intensity). This would result in measurable decorrelation of the total emission between 353 GHz and 217 GHz in pixels of low  217 GHz polarized intensity. This is an especially worrisome possibility because \texttt{target} pixels are selected for their misaligned LOS magnetic field structure, and are thus expected to have systematically lower dust polarized intensity in both frequencies.  
This is indeed the case, as demonstrated by histograms of the polarized intensities at 353 and 217~GHz in Fig.~\ref{fig:cmb_cont}.
To exclude the possibility of detecting CMB-induced frequency decorrelation and incorrectly attributing it to frequency decorrelation induced by misaligned magnetic fields in distinct dust components, we perform our analysis on maps from which the CMB contribution has been subtracted. To control against differences between component separation algorithms, we repeat the analysis on maps obtained using four different algorithms: \texttt{commander}, \texttt{nilc}, \texttt{sevem}, and \texttt{smica} (\citealt{PlaXII2014}; \citealt{PlaIX2016}; \citealt{PlaIV2018}).

The two remaining effects (sampling uncertainties and data noise) are statistical, and we deal with them through the formulation and statistical testing of two null hypotheses, discussed below. Both null hypotheses express the same physical conclusion: no LOS frequency decorrelation is detectable in {\it Planck} data. 
Rejection of these null hypotheses, consistent across different maps and implementations of the \texttt{target} sample,
will constitute evidence for the presence of frequency decorrelation induced by multiple dust components permeated by misaligned magnetic fields along selected lines of sight.

We quantify the per-pixel multi-frequency data noise by propagating the observational uncertainties on the individual Stokes parameters at the two frequencies to the measurement of the EVPA difference ($\Delta_s(\psi_{353}^i,\psi_{217}^i)$).
For pixel $i$ we thus define the multi-frequency data noise as
\begin{equation}
    {\sigma_{\Delta_s}}^i \equiv S(\{{\Delta_s(\psi_{353}^i,\psi_{217}^i)}\})
\label{eq:sigDelta_s}
\end{equation}
where the ensemble $\{{\Delta_s(\psi_{353}^i,\psi_{217}^i)}\}$ is obtained through the computation of EVPA difference on 10,000 MC simulations of noise-correlated Stokes parameters at each frequency.
Therefore in computing Eq.~\ref{eq:sigDelta_s}, the sum in Eq.~\ref{eq:CircStd} is over realizations, rather than over sample pixels as Eq.~\ref{eq:D}.

\subsection{Null hypotheses}\label{hypotheses}

\begin{figure}
\begin{center}
\begin{tabular}{c}
\texttt{control} \\[-.5ex]
\rotatebox{90}{{\large \hspace{2.9cm} ${\sigma_{\Delta_s}}^i$}}
\includegraphics[trim={1.1cm 1.2cm .6cm 1.3cm},clip,width=.95\columnwidth]{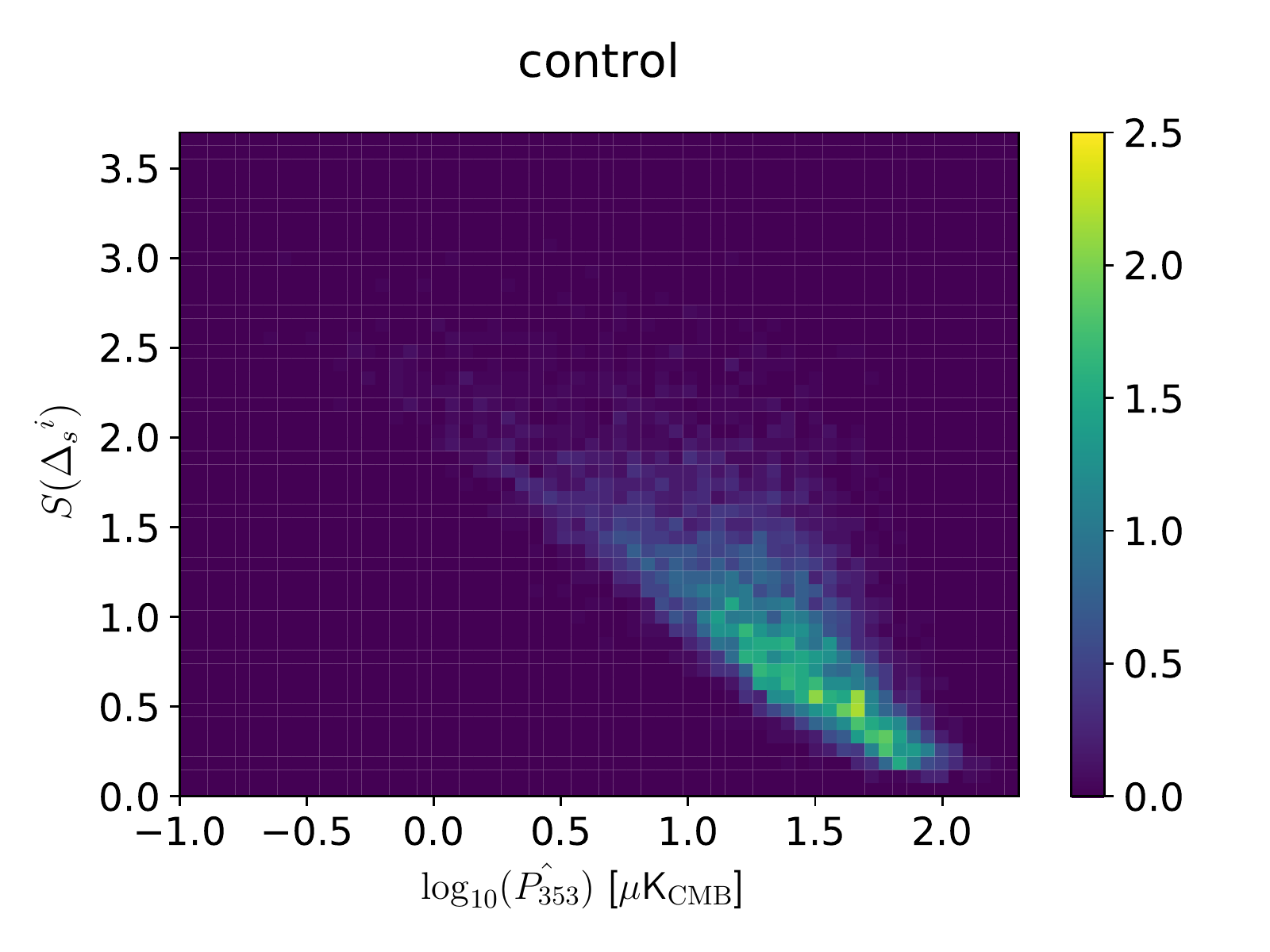}\\
{$\log_{10}(\hat{P}_{353})$ [$\mu$K$_{\rm{CMB}}$]}
\\[-1.ex] \\
\texttt{target1} \\[-.4ex]
\rotatebox{90}{{\large \hspace{2.9cm} ${\sigma_{\Delta_s}}^i$}}
\includegraphics[trim={1.1cm 1.2cm .6cm 1.3cm},clip,width=.95\columnwidth]{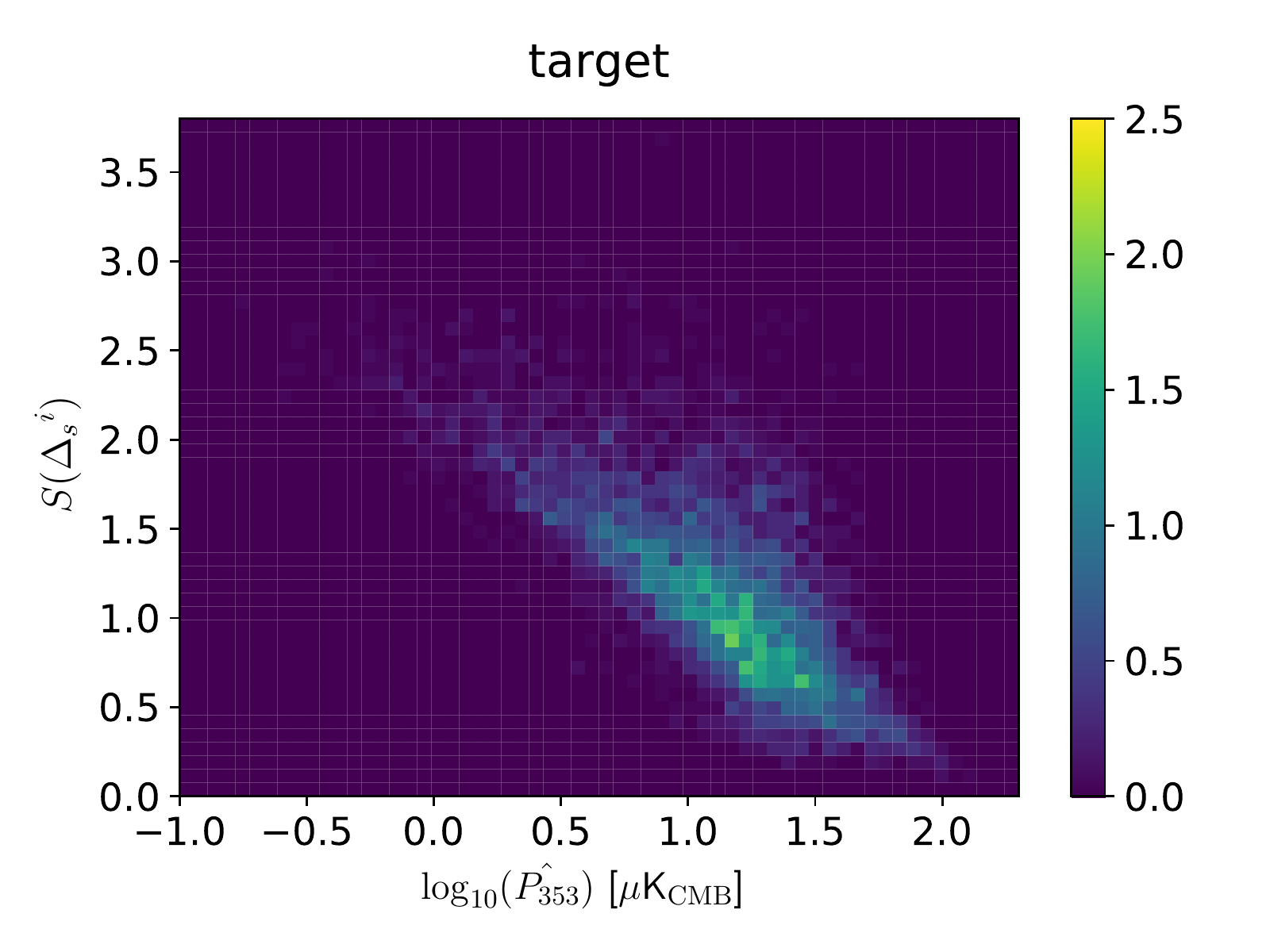}\\
{$\log_{10}(\hat{P}_{353})$ [$\mu$K$_{\rm{CMB}}$]}
\end{tabular}
\caption{Two-dimensional normalized histograms of the uncertainties in EVPA differences (${\sigma_{\Delta_s}}^i$) and debiased polarized intensity at 353 GHz ($\hat{P}_{353}$) for the \texttt{control} sample (top) and the \texttt{target1} sample (bottom), using PR3 maps, with no CMB subtraction. Both histograms are normalized and bounded to the same color scale. The two quantities are correlated: \texttt{target1} has noisier EVPA differences than \texttt{control}, because of the lower polarized intensities in its pixels. 
}
\label{fig:stdDPPAs-P3532D}
\end{center}
\end{figure}

\begin{figure}
\begin{center}
\begin{tabular}{c}
\hspace{.5cm} Sample polarization fraction and noise
\\[-.2ex]
\rotatebox{90}{\hspace{.4cm} Normalized Distribution \hspace{-1.6cm}}
\includegraphics[trim={4.2cm 4.6cm -.0cm 3.4cm},clip,width=.95\columnwidth]{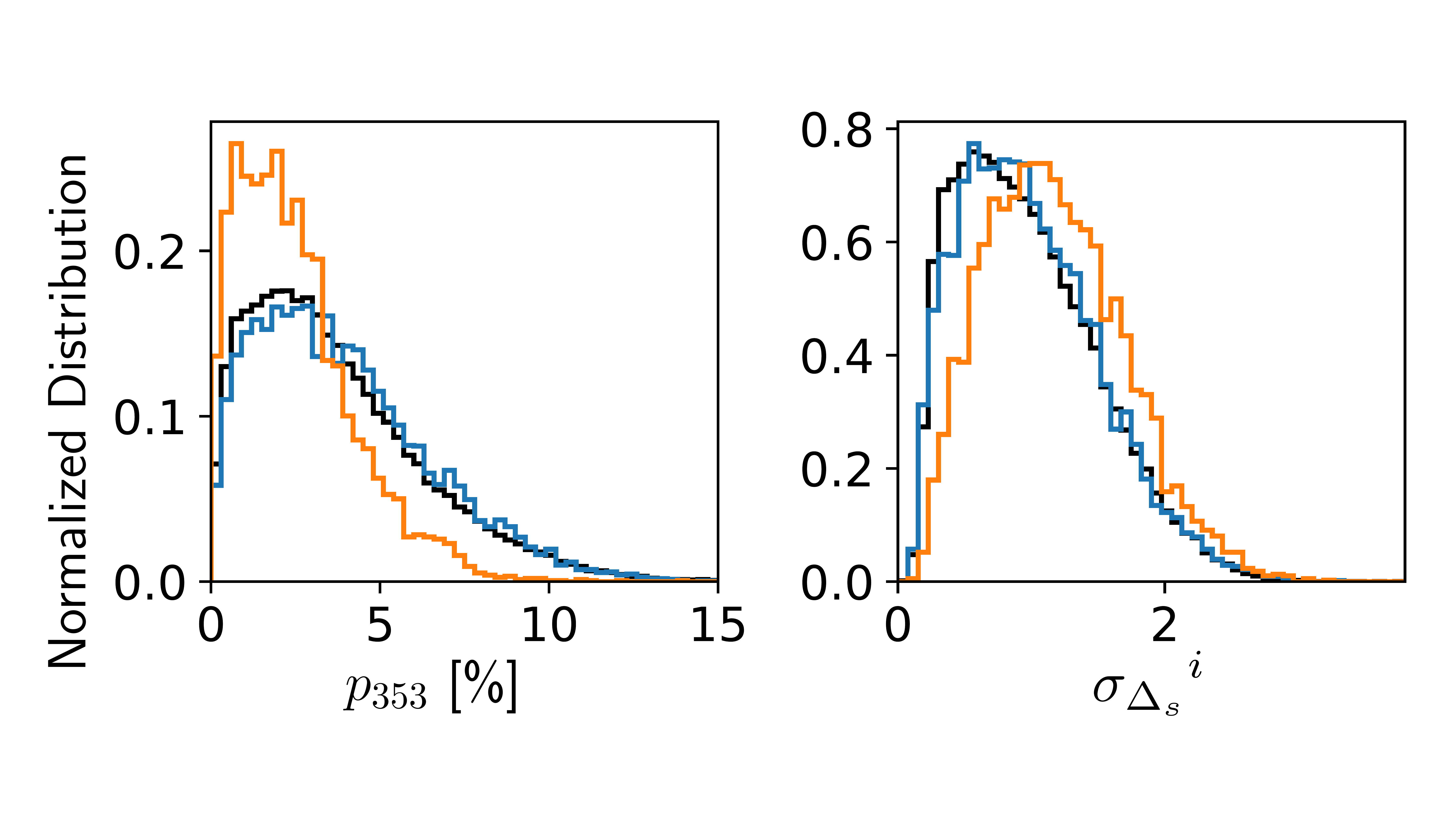}
\end{tabular}
\caption{
Histograms of polarization fraction at 353 GHz (left) and per-pixel inter-frequency uncertainty (${\sigma_{\Delta_s}}^i$, Eq.~\ref{eq:sigDelta_s})
(right), for \texttt{all} (black), \texttt{control} (blue) and \texttt{target1} (orange). Histograms correspond to PR3 polarization maps with no CMB subtraction. The \texttt{target1} sample is distinctly less polarized (left) and  noisier (right) than \texttt{all} and \texttt{control}.
}
\label{fig:pfraction_noise}
\end{center}
\end{figure}

{\bf Null Hypothesis I:} ``$\mathcal{D}_{\texttt{target}} - \mathcal{D}_{\texttt{control}} \leq 0$.'' 
The selection of the \texttt{target} and \texttt{control} samples is astrophysical and ``agnostic'' to other sources of frequency decorrelation. Once the CMB is subtracted, residual systematics are corrected, and sample size is accounted for, any significant difference in $\mathcal{D}$ between the two samples should therefore have an astrophysical explanation. There are two astrophysical reasons why $\mathcal{D}$ would differ in these samples. First, LOS frequency decorrelation (the effect we are looking for) induces an EVPA change between 217 and 353 GHz in \texttt{target}. This directly increases $\mathcal{D}$ in \texttt{target} compared to \texttt{control}. Second, LOS frequency decorrelation results in depolarization in \texttt{target} pixels.
This increases $\mathcal{D}$  {\em indirectly} in \texttt{target} compared to \texttt{control}, since a lower polarization fraction leads to a lower polarized intensity and thus a higher level of noise (e.g. see Fig.~\ref{fig:stdDPPAs-P3532D}). This difference is {\em also} attributable to the effect we are looking for.
The fact that \texttt{target} pixels are more depolarized than \texttt{control} pixels reflects the anti-correlation between $\mathcal{N}_c$ and $p_{353}$ already found in \cite{Pan2020}.
The dissimilarity of $p_{353}$ in the two samples is shown in the left panel of Fig.~\ref{fig:pfraction_noise}.
The misalignment criterion used to select \texttt{target} pixels means that these lines of sight experience more LOS depolarization. The polarized intensity and multi-frequency polarization angle uncertainty are anti-correlated (Fig.~\ref{fig:stdDPPAs-P3532D}). Thus the preferentially depolarized \texttt{target} pixels have systematically higher $\sigma_{\Delta_s}^i$ (Fig.~\ref{fig:pfraction_noise}).
We have confirmed that there is no systematic difference in the distribution of total intensity between the \texttt{target} and \texttt{control} samples at either frequency.

Therefore, we conclude that, once we have accounted for sample size, any deviation of $\mathcal{D}_{\texttt{target}} - \mathcal{D}_{\texttt{control}}$ from zero that persists across all PR3/\texttt{SRoll2} CMB-subtracted maps should be astrophysical in origin; if the direction of such a deviation is  $\mathcal{D}_{\texttt{target}} - \mathcal{D}_{\texttt{control}} > 0$, this would constitute evidence for LOS frequency decorrelation. In practice, we will calculate and report: the best-guess value $\mathcal{D}_{\texttt{target}} - \mathcal{D}_{\texttt{control}}$; its uncertainty, calculated from the individual uncertainties in $\mathcal{D}_{\texttt{target}}$ and $\mathcal{D}_{\texttt{control}}$; the p-value of the null hypothesis, $\mathcal{D}_{\texttt{target}} - \mathcal{D}_{\texttt{control}} \leq 0$. If we found the p-value to be improbably low, this would reject null Hypothesis I and constitute evidence for LOS frequency decorrelation (from a combination of depolarization and direct EVPA change) caused by misaligned magnetic fields in distinct dust components.

\begin{figure}
\begin{center}
\begin{tabular}{cc}
\rotatebox{90}{\hspace{1.2cm} {Normalized Distribution}}	&	\includegraphics[trim={1.2cm 1.3cm .2cm -.4cm},clip,height=6.3cm]{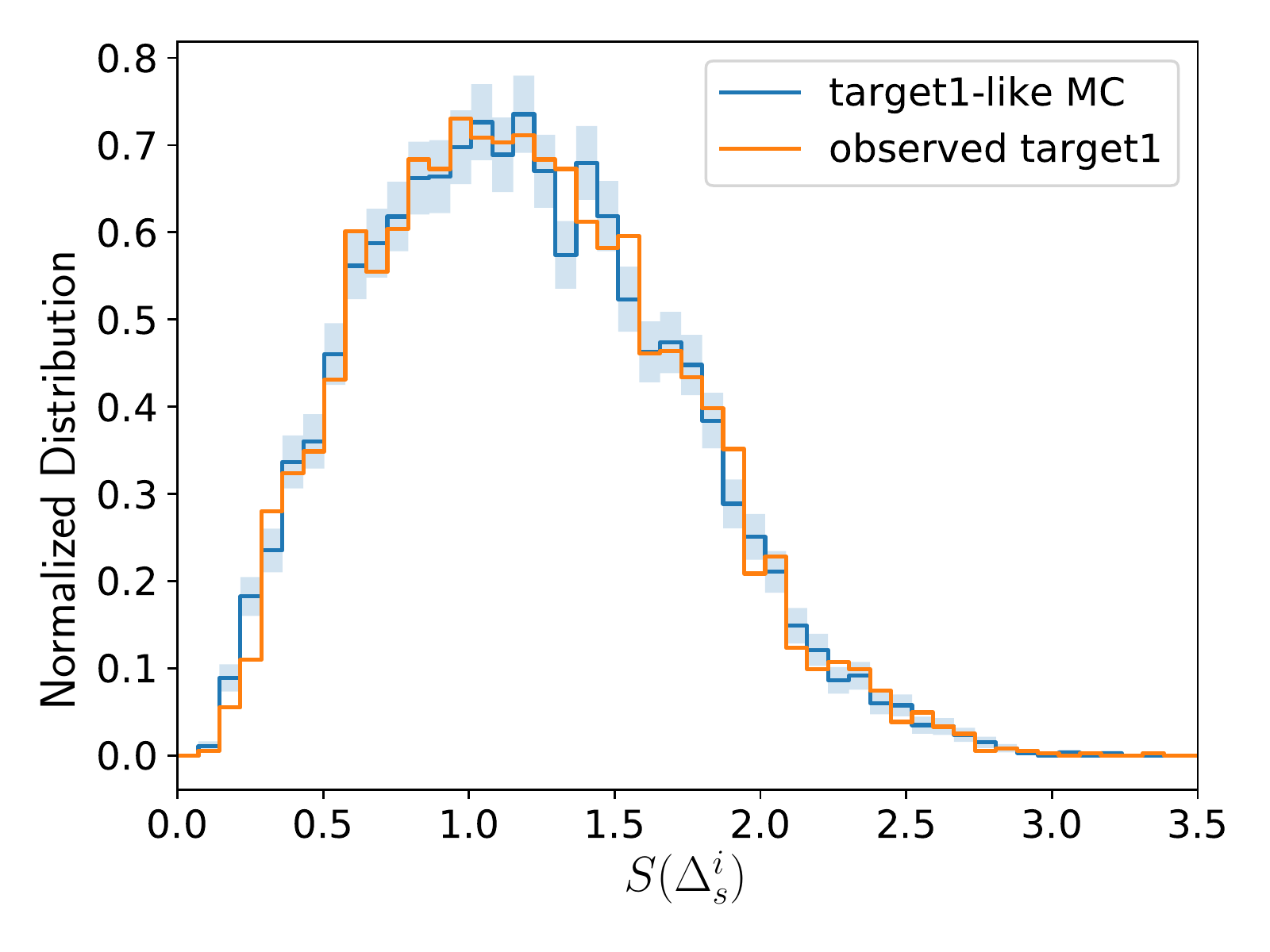} \\
   &   ${\sigma_{\Delta_s}}^i$
\end{tabular}
\caption{Effectiveness of weighted resampling in producing \texttt{target-like MC} draws from \texttt{control} with noise properties matched to \texttt{target}: means and standard deviations per bin of normalized histograms of ${\sigma_{\Delta_s}}^i$ for \texttt{target1-like MC} samples (blue), overplotted on the distribution of those uncertainties for the observed \texttt{target1} sample (orange).
The shaded blue area marks the plus and minus one standard deviation around the mean calculated in each bin from 10,000 \texttt{target1-like MC} draws obtained through weighted bootstrap resampling of \texttt{control}. They correspond to sampling uncertainties. The continuous blue line marks the mean in each bin.
Very similar results are obtained for the \texttt{target2} sample and for all combinations of set of polarization maps and CMB estimates. }
\label{fig:stdDPPAs_wB}
\end{center}
\end{figure}

{\bf Null Hypothesis II:} ``The observed \texttt{target} sample is a coincidental high-noise draw from the same parent sample as \texttt{control}.'' 
The physical consequence of this hypothesis is that any excess of $\mathcal{D}_{\texttt{target}}$ over $\mathcal{D}_{\texttt{control}}$ is entirely due to \texttt{target} being smaller and noisier\footnote{i.e., having pixels featuring larger uncertainties in $\Delta_S(\psi_{353}^i,\psi_{217}^i)$} than \texttt{control} (see Fig.~\ref{fig:stdDPPAs-P3532D} and Fig.~\ref{fig:pfraction_noise}); any direct EVPA change between 217 and 353 GHz because of LOS magnetic-field misalignment is below the noise level of {\it Planck} data. To test this hypothesis, we will generate draws from \texttt{control} that are as small and as noisy as \texttt{target}, and we will compare them with the observed \texttt{target}, using the $\mathcal{D}$ test statistic. Clearly, these "\texttt{target}-like" Monte-Carlo-generated draws will not include {\em any} EVPA change between 353 and 217 GHz due to LOS-frequency decorrelation, since all \texttt{control} pixels feature only a single cloud along that line of sight. To match the noise properties of \texttt{target}, we weight the probability of choosing a specific pixel $j$ by its value of ${\sigma_{\Delta_s}}^j$, according to the distribution of $\{{\sigma_{\Delta_s}}^i\}$ in \texttt{target} (see Fig.~\ref{fig:stdDPPAs_wB}). We then construct the distribution of $\mathcal{D}$ in these simulated \texttt{target}-like LOS-decorrelation--free draws, hereafter referred to as \texttt{target-like MC}, and calculate and report the one-sided p-value of drawing the observed $\mathcal{D}_{\texttt{target}}$ from that distribution (i.e., the probability that $\mathcal{D} \geq \mathcal{D}_{\texttt{target}}$ in that distribution).
If the observed $\mathcal{D}_{\texttt{target}}$ is improbably high compared to typical values in \texttt{target-like MC} (i.e. if the p-value is improbably low), this will reject null Hypothesis II and constitute evidence for EVPA change due to LOS-induced frequency decorrelation {\em in excess} of any increased noise in highly depolarized pixels.


\begin{figure}
\begin{center}
\begin{tabular}{cc}
& \hspace{.7cm} \includegraphics[trim={.8cm 30.1cm .8cm 2.6cm},clip,width=.7\columnwidth]{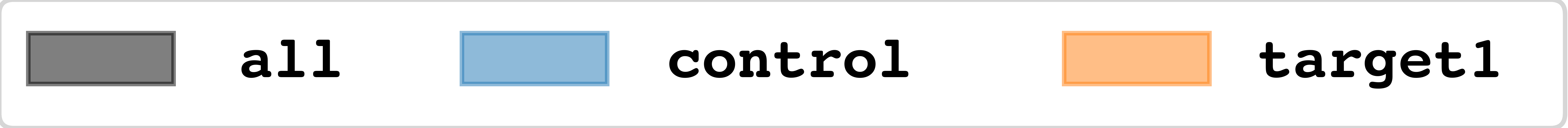}
\\[-.7ex]
\rotatebox{90}{\hspace{1.2cm} {Normalized Distribution}}	&	
\includegraphics[trim={1.1cm 1.2cm .2cm .4cm},clip,height=5.8cm]{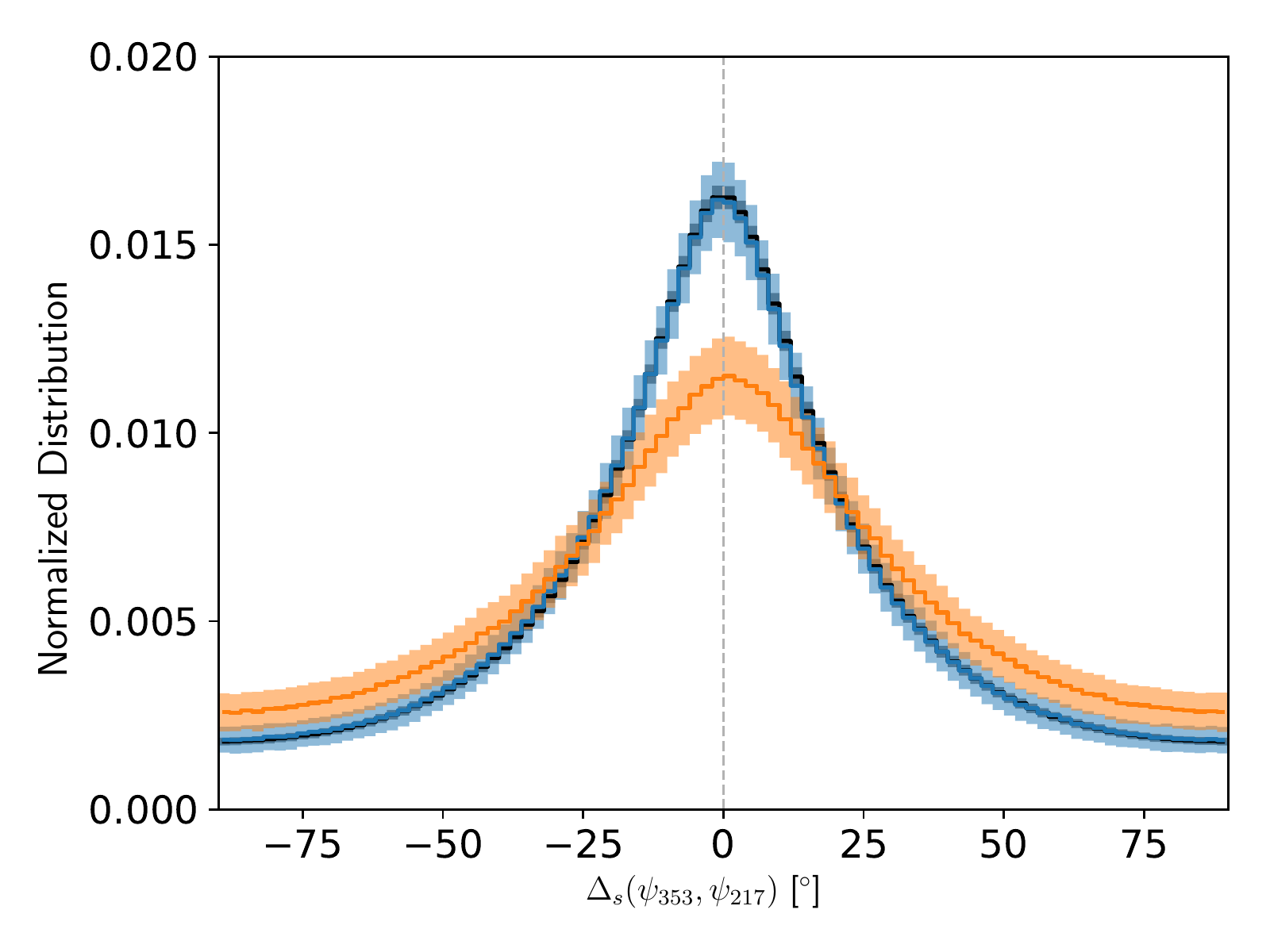}
\\
 &   {\large \hspace{.5cm} $\Delta_s(\psi_{353},\psi_{217})$}
\end{tabular}
\caption{Normalized histograms of $\Delta_s(\psi_{353},\psi_{217})$  for the \texttt{all}, \texttt{control} and \texttt{target1} samples in black, blue and orange, respectively. CMB has been subtracted from the PR3 maps using \texttt{smica}. The shaded area results from the propagation of observational uncertainties in $Q_\nu$ and $U_\nu$ down to the computation of $\Delta_s(\psi_{353},\psi_{217})$. The shaded areas mark the plus and minus one standard deviation around the means obtained in each bins of width $2^\circ$ through the MC simulations. Continuous lines show the means of the three samples.}
\label{fig:DPPAs}
\end{center}
\end{figure}

\section{Detection of LOS Frequency Decorrelation}
\label{sec:results}
Figure~\ref{fig:DPPAs} shows the normalized distributions of $\Delta_s(\psi_{353},\psi_{217})$, the signed difference of EVPAs between 353 and 217 GHz frequency bands, for the \texttt{all}, \texttt{control}, and \texttt{target1} samples in CMB-subtracted PR3 polarization maps.
The distributions for \texttt{control} and \texttt{all} are similar, while the distribution for \texttt{target1} differs noticeably from the other two by being much less peaked around zero and much more spread out.

\begin{figure*}
\begin{center}
\begin{tabular}{c}
\includegraphics[trim={3cm 7.cm 3cm 7cm},clip,height=5.8cm]{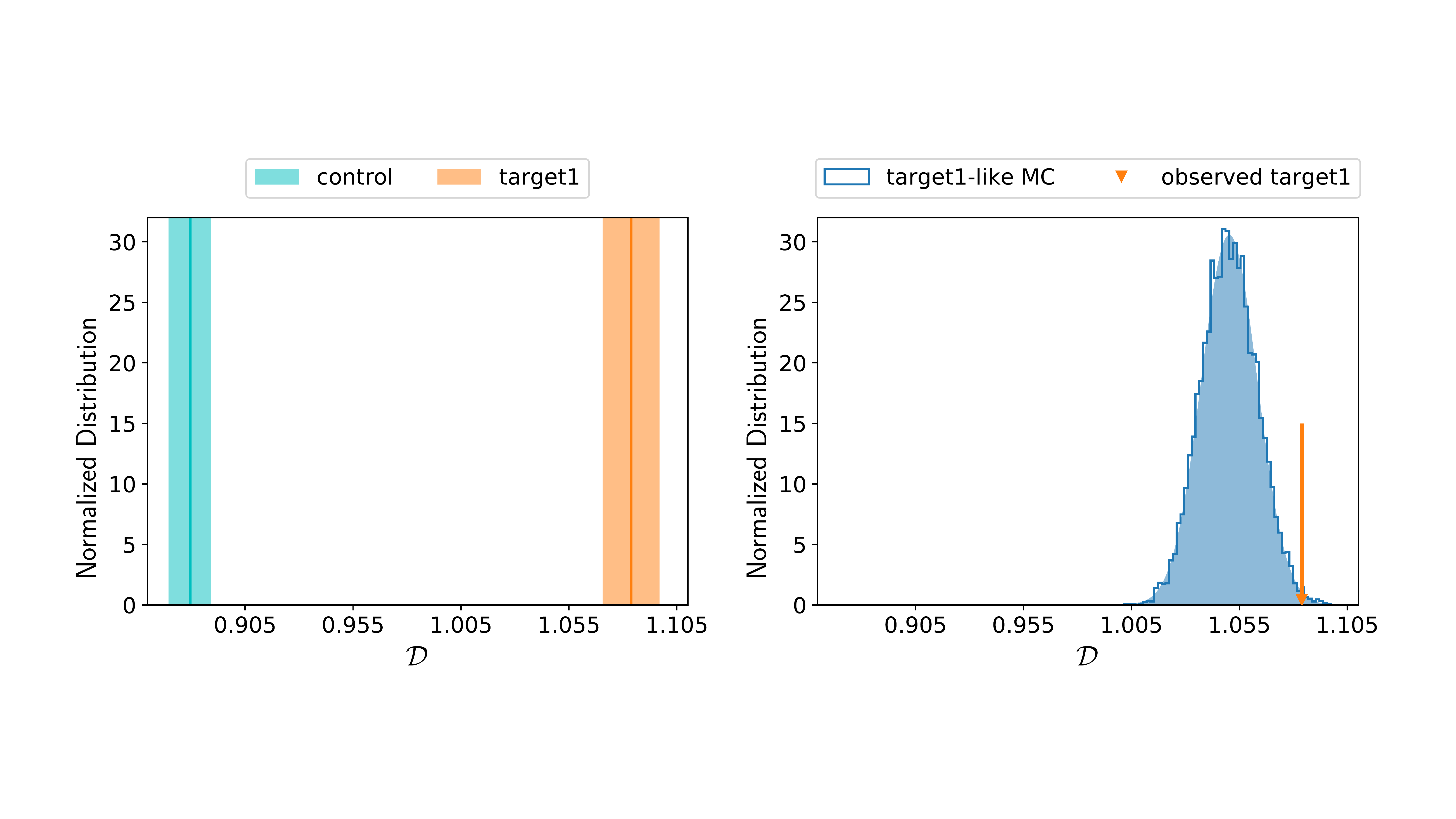}
\end{tabular}
\caption{
(Left) Rejecting Null Hypothesis I. Summary statistics of $\mathcal{D}$ values obtained through 10,000 bootstrap resampling of \texttt{control} (cyan) and \texttt{target1} (orange) samples.
The means and one standard deviations are reprensented by the thick vertical lines and shaded area respectively.
(Right) Rejecting Null Hypothesis II. The blue histogram shows the distribution of $\mathcal{D}$ values obtained through 10,000 resampling of \texttt{control} with weights that guarantee the same level of EVPA difference uncertainties in the resampled samples than in \texttt{target1}. The shaded blue distribution is a Gaussian fit to the histogram. The vertical orange arrow indicates the $\mathcal{D}$ value computed for the observed full \texttt{target1} sample.
The examples shown in both panels make use of the PR3 polarization maps from which we have subtracted the smica CMB estimate.
Results are consistent with all other implementations as shown in Tables~\ref{tab:HypoI} and~\ref{tab:SL_F3}.
}
\label{fig:D_wBDPPAs}
\end{center}
\end{figure*}

To test Null Hypothesis I, we calculate $\mathcal{D}$, the spread of the distribution of EVPA differences, for \texttt{target} and \texttt{control}, for both implementations of \texttt{target}, both sets of {\it Planck} polarization maps, and all CMB estimates from the four component separation algorithms.
We also calculate uncertainties of $\mathcal{D}$ through unweighted bootstrapping for each of these cases. The left panel of Fig.~\ref{fig:D_wBDPPAs} shows $\mathcal{D}_{\texttt{control}}$ and $\mathcal{D}_{\texttt{target1}}$, with their respective uncertainties, for the PR3 map, from which the \texttt{smica} CMB estimate has been subtracted.
It is obvious that $\mathcal{D_{\texttt{target1}}}$ is very significantly larger than $\mathcal{D_{\texttt{control}}}$, so we expect that Null Hypothesis I is rejected at very high significance, providing clear evidence for the presence of LOS frequency decorrelation in {\it Planck} data.
Since the distributions of the $\mathcal{D}_{\texttt{control}}$ and $\mathcal{D}_{\texttt{target}}$ obtained from the bootstrapped samples are very nearly Gaussian, the mean of their difference will be the difference of their means, and the uncertainty of their difference can be obtained from their individual uncertainties added in quadrature. 
These values are given in Table~\ref{tab:HypoI}, together with the one-sided p-value of Null Hypothesis I ("$\mathcal{D}_{\texttt{target}}-\mathcal{D}_{\texttt{control}}\leq 0$"). Indeed, Null Hypothesis I is very strongly rejected for both sets of maps (PR3 vs \texttt{SRoll2}), all CMB subtraction algorithms, and both \texttt{target} implementations. 
\begin{table*}
\caption{Testing Null Hypothesis I. Probability distribution of the difference of $\mathcal{D}$ values computed for \texttt{control} and \texttt{target}, with their uncertainties, computed from the sampling uncertainties in $\mathcal{D}_{\texttt{target}}$ and $\mathcal{D}_{\texttt{control}}$, in turn obtained through unweighted bootstrapping. The one-sided p-value gives the probability that $\mathcal{D}_{\texttt{target}} \leq \mathcal{D}_{\texttt{control}}$.
Results are given for both for PR3 and \texttt{SRoll2} polarization maps; for removal of  the CMB polarization as estimated from the different Legacy component separation methods, as well as for no CMB removal; and for our two implementations of the \texttt{target} pixel selection, presented in Sect.~\ref{subsec:pixelsamples}. }
\label{tab:HypoI}
\centering
\begin{tabular}{l cc | cc || cc | cc}
\hline\hline
\\[-1.5ex] CMB Removal & \multicolumn{4}{c}{Implementation 1} & \multicolumn{4}{c}{Implementation 2} \\
\\[-1.5ex]
 & \multicolumn{2}{c}{PR3} & \multicolumn{2}{c}{\texttt{SRoll2}} & \multicolumn{2}{c}{PR3} & \multicolumn{2}{c}{\texttt{SRoll2}} \\
\\[-1.5ex]
\hline \\[-1.5ex]
    & diff. & p-value &     diff. & p-value & diff. & p-value & diff. & p-value \\
\\[-.5ex]
\hline
\\[-.5ex]
None                & $0.22\pm0.02$  &  $7 \times 10^{-34}$  &  $0.28\pm0.02$  &  $2  \times  10^{-48}$	& $0.19\pm0.017$ & $4  \times 10^{-29}$   &   $0.25\pm0.018$ & $6  \times 10^{-44}$ \\
\texttt{commander}  & $0.20\pm0.02$  &  $7  \times 10^{-34}$  & $0.24\pm0.02$  &  $4  \times 10^{-48}$	& $0.17\pm0.015$ & $1  \times 10^{-28}$   &   $0.21\pm0.016$ & $1  \times 10^{-40}$ \\
\texttt{nilc}       & $0.20\pm0.02$  &  $2  \times 10^{-35}$  & $0.26\pm0.02$  &  $2  \times 10^{-54}$	& $0.18\pm0.015$ & $4  \times 10^{-32}$   &   $0.23\pm0.016$ & $6  \times 10^{-49}$ \\
\texttt{sevem}      & $0.19\pm0.02$  &  $3  \times 10^{-33}$  & $0.24\pm0.02$  &  $7 \times  10^{-46}$	& $0.17\pm0.015$ & $2  \times 10^{-29}$   &   $0.21\pm0.016$ & $1  \times 10^{-40}$ \\
\texttt{smica}      & $0.20\pm0.02$  &  $4  \times 10^{-36}$  &  $0.25\pm0.02$  &  $3  \times 10^{-49}$	& $0.18\pm0.015$ & $9  \times 10^{-32}$   &   $0.22\pm0.016$ & $5  \times 10^{-45}$ \\
\\[-1.5ex]
\hline
\end{tabular}
\end{table*}
\begin{table*}
\caption{Testing Null Hypothesis II. Summary statistics of the $\mathcal{D}$ values computed for weighted subsamples of \texttt{control} with level of EVPA difference uncertainties (${\sigma_{\Delta_s}}^i$) matching those of \texttt{target} and of size equal to the size of \texttt{target} (referred to as \texttt{target-like MC}), compared to the observed $\mathcal{D}$ value of \texttt{target}.
The probability of $\mathcal{D}_{\texttt{target}}$ to arise as a random realization of a ${\sigma_{\Delta_s}}^i$-matched \texttt{control} subsample of size equal to the size of \texttt{target} is also quantified in terms of a p-value for each case studied. The information is presented both for PR3 and \texttt{SRoll2} polarization maps and when removing the CMB polarization as estimated from the different Legacy component separation methods. Results are shown for both implementations of the \texttt{target} pixel selection presented in Sect.~\ref{subsec:pixelsamples}. We also provide the results for the case of no CMB removal.
}
\label{tab:SL_F3}
\centering
\begin{tabular}{l l | c c || c c}
\hline\hline
\\[-1.5ex] CMB Removal & & \multicolumn{2}{c}{Implementation 1} & \multicolumn{2}{c}{Implementation 2} \\
\\[-1.5ex]
 & & PR3 & \texttt{SRoll2} & PR3 & \texttt{SRoll2} \\
\\[-1.5ex]
\hline \\[-1.5ex]
\multirow{3}{*}{None}
    &	$\mathcal{D}_{\texttt{target-like MC}}$	& $1.161\pm0.014$	& $1.182\pm0.015$	& $1.152\pm0.013$	& $1.174\pm0.013$\\
    &	$\mathcal{D}_{\texttt{target}}$	& $1.224$ & $1.283$	& $1.195$	& $1.253$\\
\\[-1.5ex]
   &  p-value 					& $5\times 10^{-6}$		& $4\times 10^{-12}$	& $6\times 10^{-4}$		& $4\times 10^{-9}$	\\
\\[-1.5ex]
\hline \\[-1.5ex]
\multirow{3}{*}{\texttt{commander}}
    &	$\mathcal{D}_{\texttt{target-like MC}}$	& $1.036\pm0.013$ & $1.069\pm0.013$	& $1.019\pm0.012$	& $1.047\pm0.012$\\
    &	$\mathcal{D}_{\texttt{target}}$	& $1.067$	& $1.113$	& $1.040$	& $1.084$\\
\\[-1.5ex]
    &   p-value					& $7 \times 10^{-3} $	& $4\times10^{-4}$	    & $4\times 10^{-2}$ & $10^{-3}$		\\
\\[-1.5ex]
\hline \\[-1.5ex]
\multirow{3}{*}{\texttt{nilc}}
    &	$\mathcal{D}_{\texttt{target-like MC}}$	& $1.031\pm0.013$	& $ 1.063\pm0.013$ & $1.014\pm0.012$	& $1.043\pm0.012$	\\
    &	$\mathcal{D}_{\texttt{target}}$	& $1.071$	& $ 1.128$ & $1.047$	& $1.100$	\\
\\[-1.5ex]
    &   p-value						& $10^{-3}$ & $4\times10^{-7}$     & $3\times 10^{-3}$ &$2\times 10^{-6}$  		\\
\\[-1.5ex]
\hline \\[-1.5ex]
\multirow{3}{*}{\texttt{sevem}}
    &	$\mathcal{D}_{\texttt{target-like MC}}$	& $1.024\pm0.013$	& $1.056\pm0.013$	& $1.010\pm0.012$	& $1.037\pm0.012$\\
    &	$\mathcal{D}_{\texttt{target}}$	& $1.052$	& $1.095$	& $1.029$	& $1.069$\\
\\[-1.5ex]
    &   p-value 				& $2\times10^{-2} $& $2\times10^{-3}$   & $5\times 10^{-2}$		&	$4\times 10^{-3}$   \\
\\[-1.5ex]
\hline \\[-1.5ex]
\multirow{3}{*}{\texttt{smica}}
    &	$\mathcal{D}_{\texttt{target-like MC}}$	& $1.052\pm0.013$	& $1.084\pm0.014$	& $1.034\pm0.012$	& $1.037\pm0.012$\\
    &	$\mathcal{D}_{\texttt{target}}$	& $1.084$	& $1.129$	& $1.059$	& $1.106$ \\
\\[-1.5ex]
    &  p-value				& $7\times10^{-3}$&   $5\times10^{-4}$ &  $2\times 10^{-2}$		& $4\times 10^{-4}$ \\
\\[-1.5ex]
\hline
\end{tabular}
\end{table*}

\begin{figure*}
\begin{center}
\begin{tabular}{c}
\includegraphics[trim={1cm 7.cm -1cm 8cm},clip,height=5.4cm]{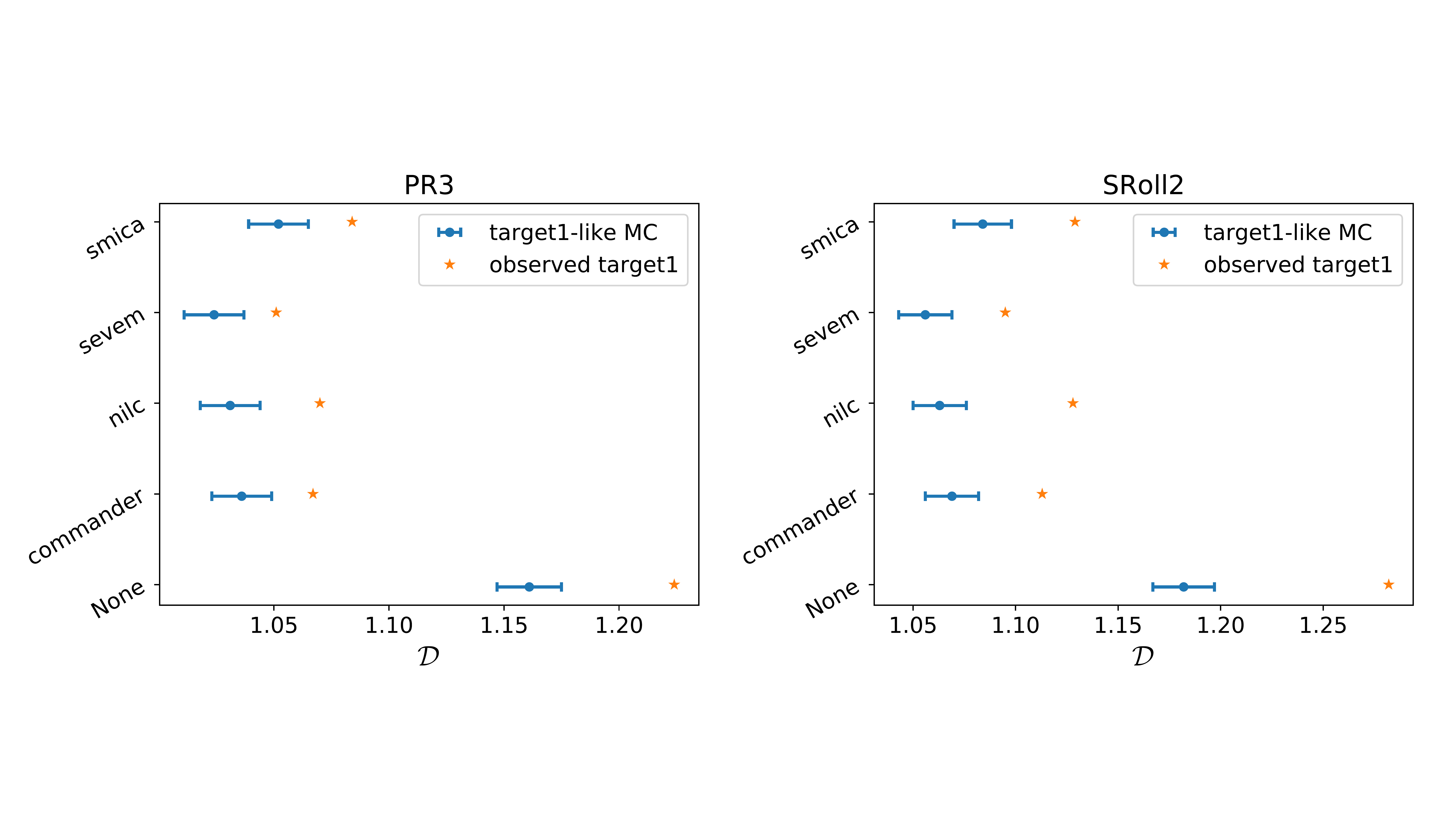}
\end{tabular}
\caption{
Summary statistics of $\mathcal{D}$ distributions for \texttt{target1}-like simulations obtained from \texttt{control} while subtracting different CMB estimates from the PR3 polarization maps (left) and \texttt{SRoll2} polarization maps (right), compared to the $\mathcal{D}$ value of \texttt{target1}. This illustrates part of the information given in Table~\ref{tab:SL_F3}.
}
\label{fig:summaryD}
\end{center}
\end{figure*}

We have thus established that \texttt{target} has statistically greater polarization angle differences between frequencies than \texttt{control},
and that this is not an effect of increased CMB contribution in \texttt{target} pixels, nor an artifact of the CMB estimate produced by any specific component separation algorithm, nor an artifact of spatially correlated residual systematics.

We now proceed to test whether this excess decorrelation is also significant {\em beyond what would be justified by the increased noise level of \texttt{target} compared to \texttt{control}} (i.e., test whether Null Hypothesis II is also rejected). For each case considered, we generate 10,000 \texttt{target-like MC} draws through noise-weighted sub-sampling from \texttt{control}, as described in the previous section; we calculate the $\mathcal{D}$ test-statistic for each; we construct the distribution of $\mathcal{D}$; and we compute the one-sided p-value that describes the probability that the $\mathcal{D}$ measured for \texttt{target} could be measured for a random pixel sample that is as small as \texttt{target}, as {\em noisy} as \texttt{target}, but completely free of LOS decorrelation according to the best current knowledge of the 3D magnetized ISM. 

One example of this process is visually represented in the right panel of Fig.~\ref{fig:D_wBDPPAs} for the case of Implementation I of \texttt{target} and PR3 maps from which the \texttt{smica} CMB estimate has been subtracted. It is clear that the observed \texttt{target} is highly decorrelated, even compared to comparably high-noise draws from \texttt{control}. Summary statistics for the $\mathcal{D}$ distributions and p-values obtained from all samples in both our implementations are reported in Table~\ref{tab:SL_F3}, while a visual representation of these results for all combinations of maps/CMB subtraction algorithms and for Implementation I of \texttt{target} is shown in Fig.~\ref{fig:summaryD}. For comparison, both in  Table~\ref{tab:SL_F3} and in Fig.~\ref{fig:summaryD}, we also provide the results of our analysis on maps {\em without} any subtraction of the CMB contribution. Indeed, the p-value of Null Hypothesis II is low for both sets of maps (PR3 vs \texttt{SRoll2}), both implementations of \texttt{target}, and all CMB estimate subtractions, with p-values ranging from $4\times 10^{-3}$ to $5\times 10^{-2}$ for PR3 maps, and from $4\times 10^{-7}$ to $4 \times 10^{-3}$ for \texttt{SRoll2} maps. 

Null Hypothesis II is systematically rejected at a higher significance for \texttt{SRoll2} maps than for PR3 maps, if all other features of the analysis remain the same. The most straightforward way to interpret this trend is that residual systematics in PR3 maps act as an additional source of noise; the \texttt{SRoll2} corrections for these systematics reduces the noise, and the LOS frequency decorrelation stands out more.
Additionally, the significance of the effect is always higher when the CMB has not been subtracted, confirming that indeed the CMB makes a distinct contribution to the difference between 353 and 217 GHz EVPAs, and that difference is more pronounced in the lower-polarized-intensity pixels of \texttt{target}. 

The robustness of the low p-value of Null Hypothesis II across maps, CMB-subtraction algorithms, and \texttt{target} selections gives us confidence that the effect is real, and that LOS frequency decorrelation due to multiple dust components is present in {\it Planck} data and detectable above the noise level -- as long as one knows where in the sky to look for it.

\section{Validation}
\label{sec:validation}
In this section we discuss additional validation tests, both statistical and physical, to increase our confidence that we have in fact detected LOS-induced frequency decorrelation in {\it Planck} data. 

\subsection{Sky distribution of \texttt{target} and \texttt{control} pixels }
Pixels of \texttt{target} and \texttt{control} sample largely disjoint parts of the sky  (see Sect.~\ref{subsec:pixelsamples}). It is thus conceivable that their difference in observed $\mathcal{D}$ might stem from different local properties, and most notably different instrumental noise or systematic properties of the data. In principle, our test of Hypothesis II, where the observed $\mathcal{D}_{\texttt{target}}$ is compared to that of noise-matched subsamples of \texttt{control}, should take into account the difference in noise properties; and the comparison between PR3 and \texttt{SRoll2} maps is performed exactly to evaluate the impact of the residual systematics.
Nevertheless, we performed two additional tests to verify that some additional, hidden, spatial correlation bias is not generating a false-positive detection of LOS frequency decorrelation. 

\begin{figure}
\begin{center}
\begin{tabular}{c}
\includegraphics[trim={0.4cm .5cm 0.4cm 1.6cm},clip,width=.98\columnwidth]{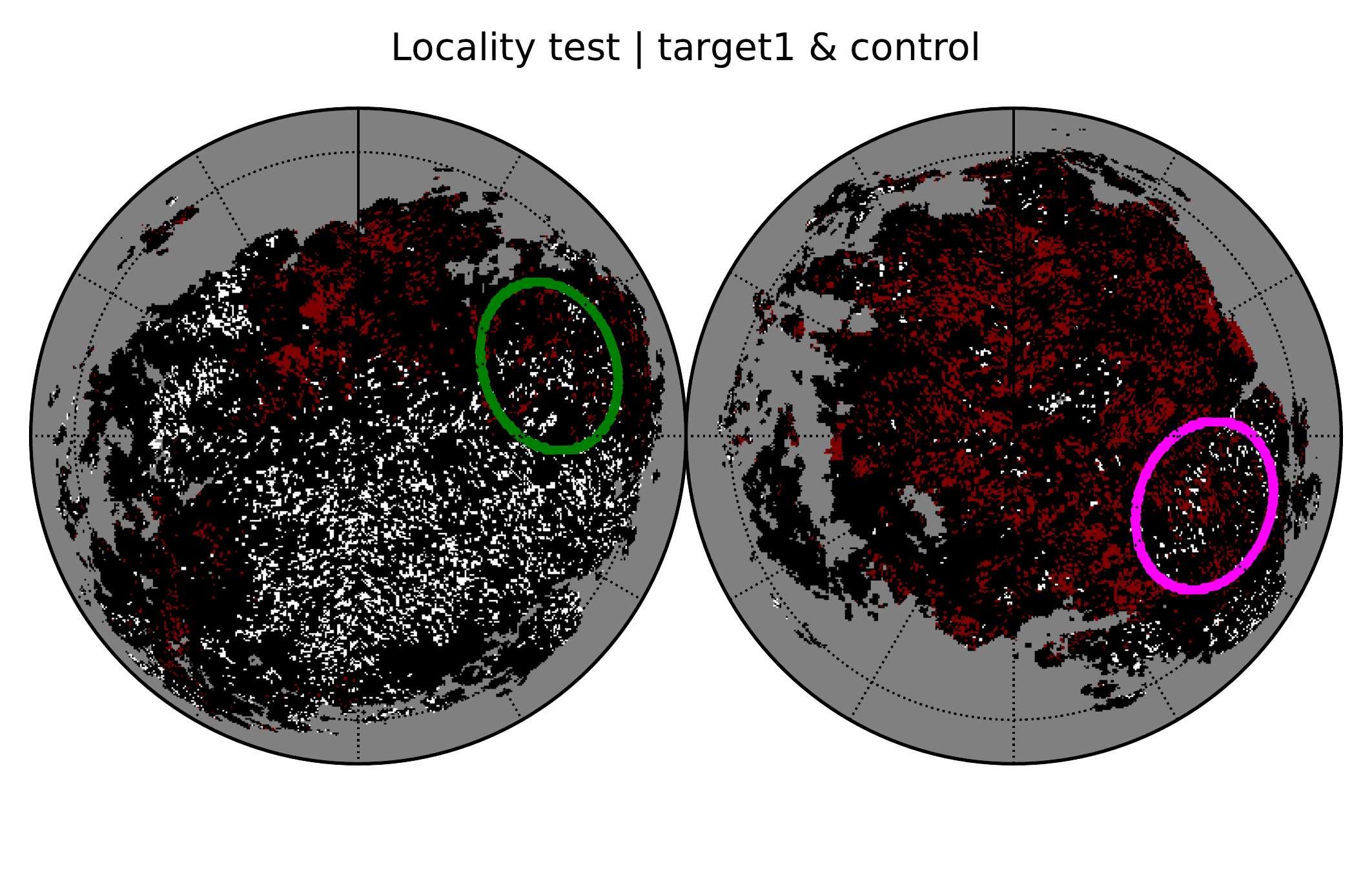}
\\
\end{tabular}
\caption{Map showing the location of sky pixels belonging to the \texttt{target1} (white) and \texttt{control} (orange) samples. Black pixels are those in \texttt{all} but neither in \texttt{target1} nor \texttt{control}. The gray area are pixels where $\mathcal{N}_c$ has not be determined (see \citealt{Pan2020}). The location of the northern and southern sky patches studied in order to investigate the effect of \texttt{target} and \texttt{control} sampling different sky regions are shown with the green and magenta circles, respectively.}
\label{fig:patches}
\end{center}
\end{figure}

\begin{figure}
\begin{center}
\begin{tabular}{cc}
{\hspace{.6cm}} Northern Patch & {\hspace{.3cm}} Southern Patch \\
[-.6ex]
\rotatebox{90}{{\hspace{.5cm}} Normalized Distribution}
\includegraphics[trim={1.1cm 1.3cm 0.4cm 1.2cm},clip,height=4.6cm]{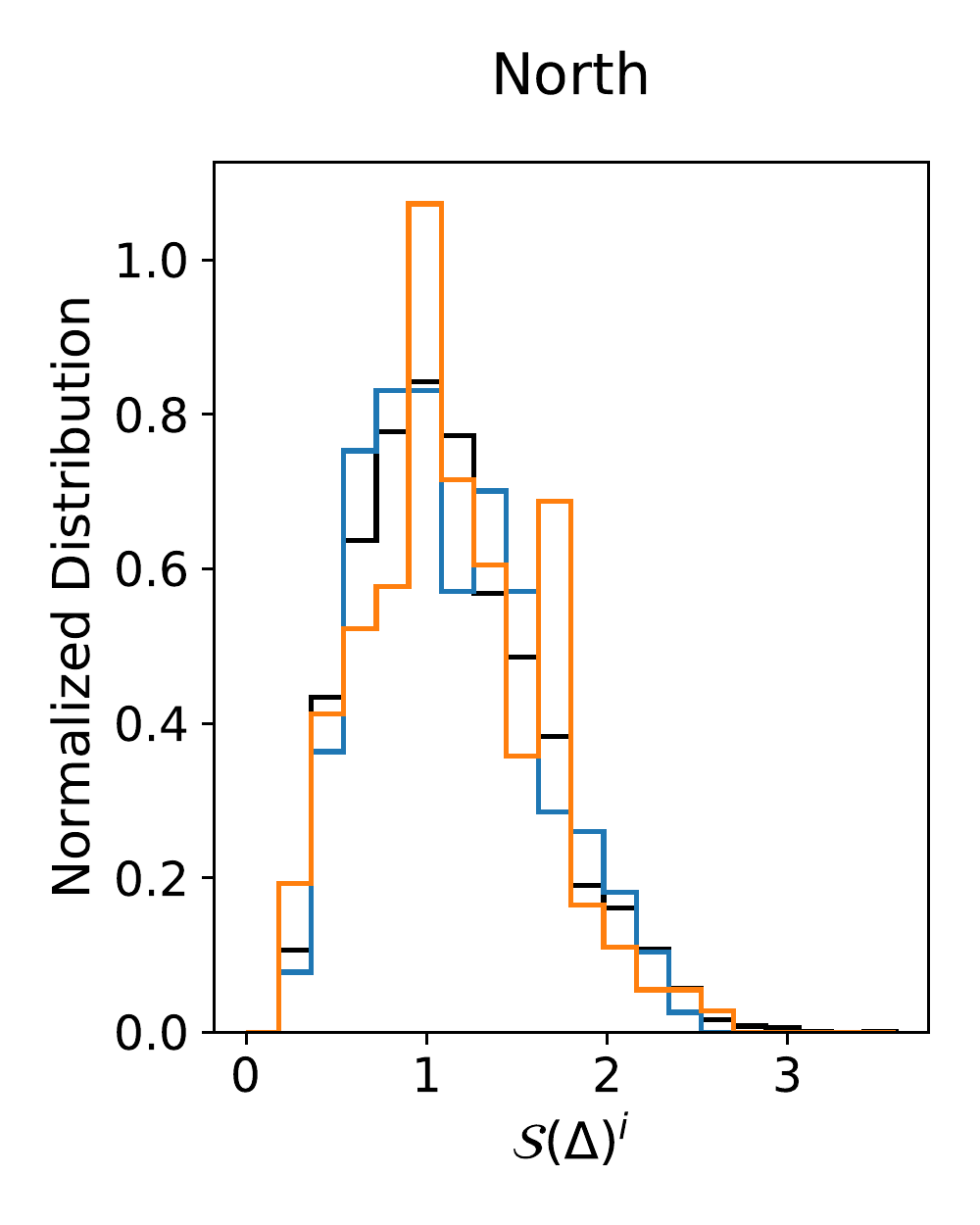}
	&
	\includegraphics[trim={1.2cm 1.3cm 0.4cm 1.2cm},clip,height=4.6cm]{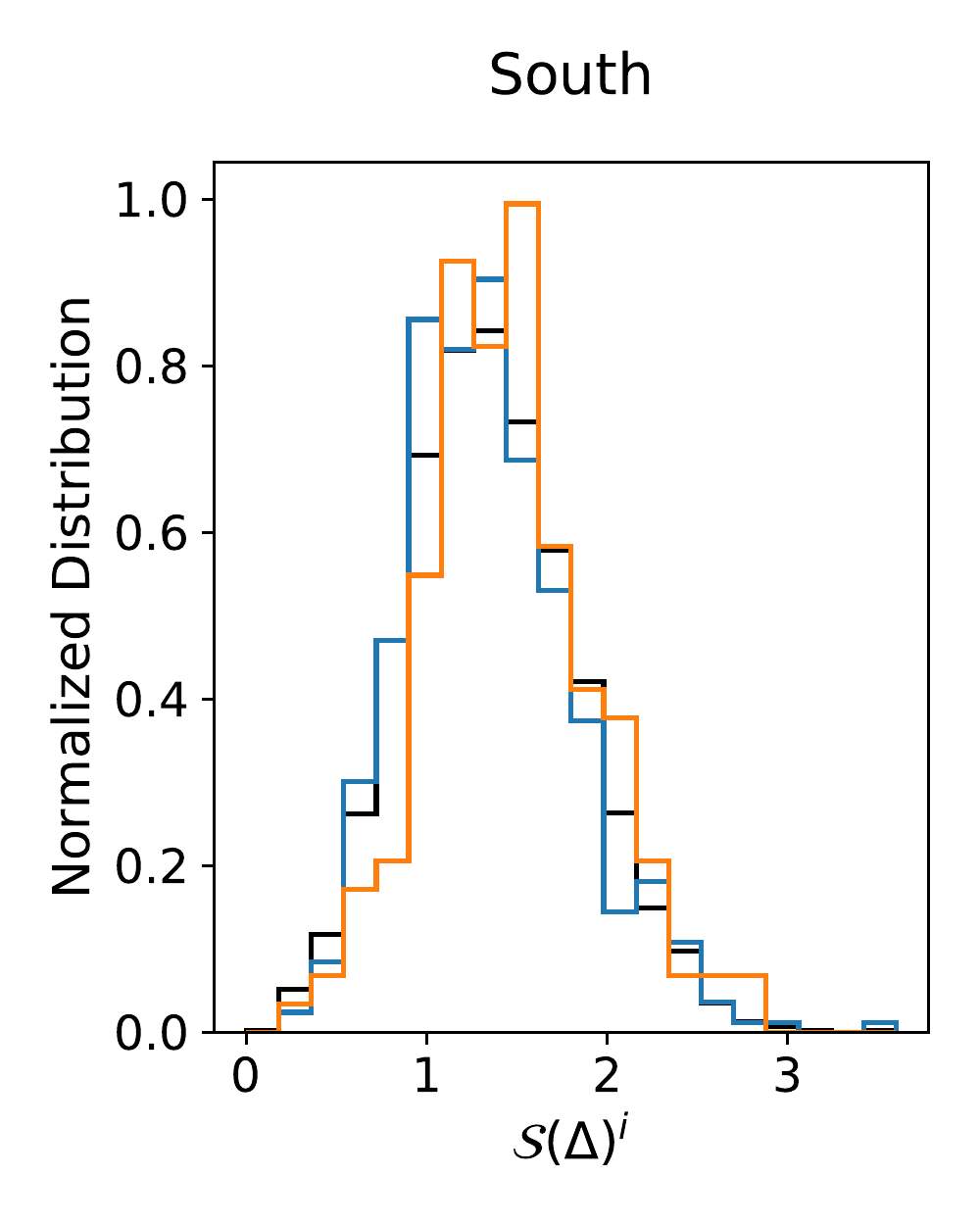}
\\[-.3ex]
{\hspace{.5cm}} ${\sigma_{\Delta_s}}^i$   &   {\hspace{.2cm}} ${\sigma_{\Delta_s}}^i$ \\
\end{tabular}
\caption{Normalized histograms of ${\sigma_{\Delta_s}}^i$ as measured on PR3 maps in the northern (left) and southern (right) sky patches for the different sub-samples of \texttt{target1} (in orange), \texttt{control} (in blue) and \texttt{all} (in black).
}
\label{fig:noise-patches}
\end{center}
\end{figure}

First, we repeated our analysis inside two sky patches that contain intermixed \texttt{target1} and \texttt{control} pixels, and that are sufficiently small so that instrumental systematics would not vary considerably within each patch. 
These sky patches were defined as regions with an angular radius of 15$^\circ$, centered on $(l,\,b)=(70^\circ,\,50^\circ)$ in the North, and $(l,\,b)=(-110^\circ,\,-50^\circ)$ in the South.
These regions were visually identified and are indicated by green and magenta outlines, respectively, in Fig.~\ref{fig:patches}. These patches as a whole contain 3352 (north) and 3353 (south) pixels. Of those, in the northern (southern) patch, 202 (162) are \texttt{target1} pixels, and 214 (461) are \texttt{control} pixels. The noise properties within each patch are overall consistent between samples (Fig.~\ref{fig:noise-patches}), unlike the full \texttt{target} and \texttt{control} samples (Fig.~\ref{fig:pfraction_noise}). We found that for all combinations of maps, CMB subtraction algorithms, and \texttt{target} sample implementations, $\mathcal{D}_{\texttt{target}}$ is larger than $\mathcal{D}_{\texttt{control}}$. The sample sizes are now too small for Null Hypothesis II to be rejected through the weighted-resampling analysis discussed in Sect.~\ref{hypotheses}; we have however verified through sub-sampling of the full \texttt{target1} and \texttt{control} samples, that the behavior of both the distribution of $\mathcal{D}_{\texttt{target-like MC}}$ and the observed $\mathcal{D}_{\texttt{target}}$ in these sky patches is consistent with what we would expect given the local noise properties and the decrease in sample size.

Second, having observed that noise properties differ systematically between northern and southern hemispheres, we repeated our analysis in the northern hemisphere alone. We chose the northern hemisphere because it contains more \texttt{target} pixels: lines of sight intersecting multiple, misaligned clouds are evidently more common in the northern Galactic sky.
We found that, despite the modest decrease in sample size for \texttt{target}, in this case the significance with which Null Hypothesis II is rejected in fact {\em increases} (p-value decreases), because in general pixels in the north are less noisy.

\subsection{Projected Rayleigh Statistic}
In order to strengthen our analysis and confirm that our results do not depend critically on our choice of $\mathcal{D}$ as our test statistic, we have repeated our analysis using the Projected Rayleigh Statistic (PRS) to quantify the degree of alignment of EVPA between frequencies.
The PRS ($Z_x$) is computed as (e.g., \citealt{Jow2018}):
\begin{equation}
    Z_x = \frac{1}{N} \sum_{i=1}^{N}{\cos{(2 \xi_i)}}
\end{equation}
where $\xi_i$ is defined in the range $\left[-\pi/2 ,\, \pi/2 \right]$, so that $Z_x$ takes values between -1 and 1.
Computing the PRS for a sample of signed difference angles $\Delta_s(\psi_{353},\psi_{217})$ defined in Eq.~\ref{eq:Delta_s}, we can quantify the level of alignment of EVPA between 353 and 217 GHz. We expect $Z_x$ to be smaller for samples with statistically larger EVPA differences.
We reproduced the analysis presented in Sect.~\ref{sec:analysis} using the PRS in place of the circular standard statistic ($S$), in order to quantify the degree of alignment/misalignment in our samples and quantitatively compare them. We found that the significance with which our hypotheses are rejected in each case are generally consistent, with no strong dependence on the choice of test statistic.

\subsection{$\mathcal{D}$ versus $\Delta(\theta_{LVC},\theta_{IVC})$}

According to the simplest two-cloud model (\citealt{Tas2015}), if the EVPA differences between frequencies are due to SED differences and magnetic field misalignment between the dust clouds, then for an ensemble of sky pixels we expect to see ({\it i}) a decrease of degree of polarization and ({\it ii}) an increase of LOS frequency decorrelation (which we quantify using $\mathcal{D}$) as $\Delta(\theta_{LVC},\theta_{IVC})$ increases.

\begin{figure}
\begin{center}
\begin{tabular}{rc}
\rotatebox{90}{\hspace{2.7cm} {\large $\mathcal{D}$}}
 & \includegraphics[trim={1.1cm 1.1cm -.4cm .0cm},clip,height=5.6cm]{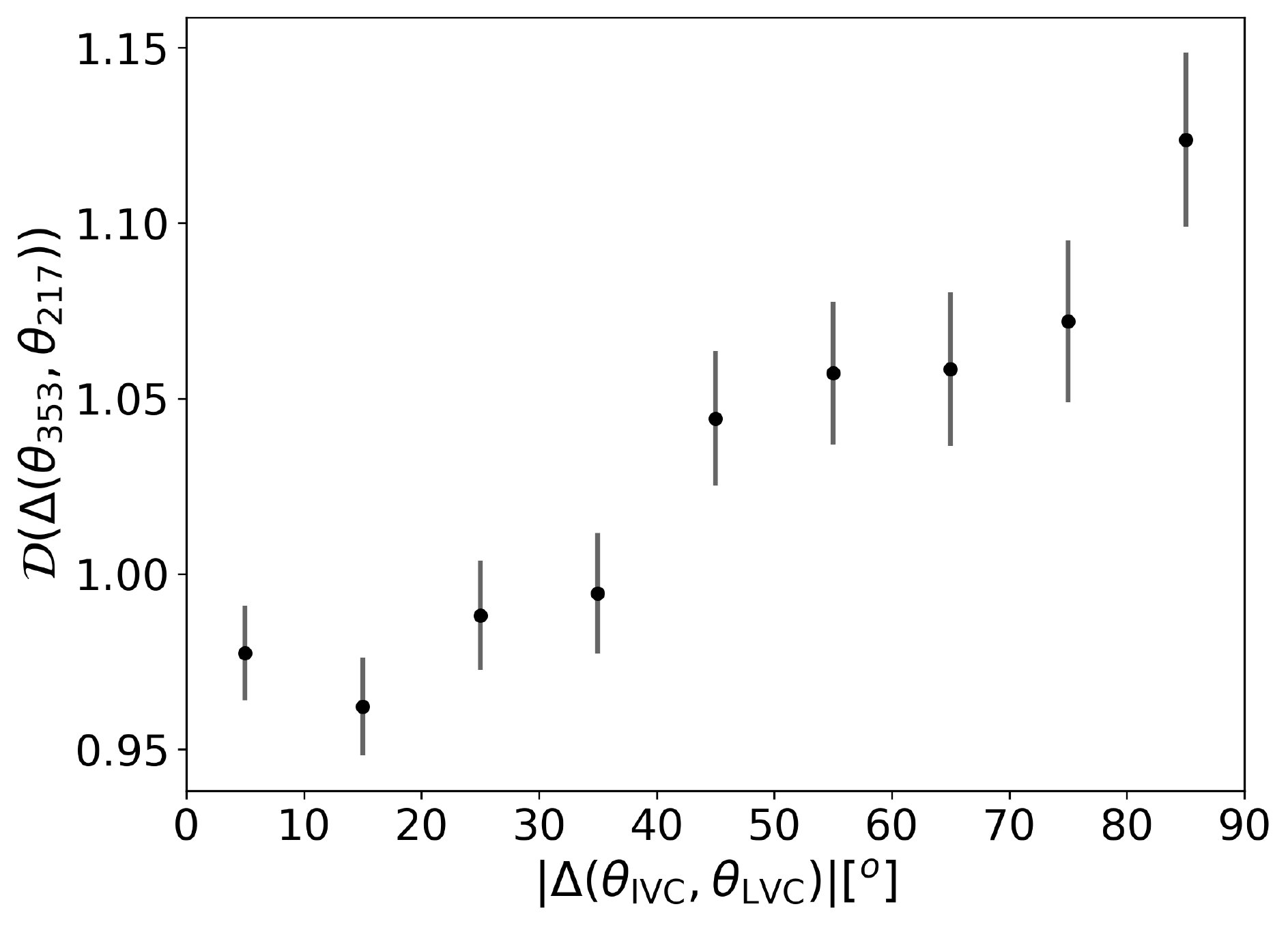} \\
 & \hspace{.7cm} {\large $\Delta(\theta_{LVC},\theta_{IVC})$ [$^\circ$]}
\end{tabular}
\caption{
Increase of the spread of EVPA differences between 353 and 217 GHz as a function of offset angle between \hi structures from integration in LVC and IVC ranges (`Implementation 1'). All sky pixels with $N_c > 1.5$ are binned according to their $\Delta(\theta_{IVC},\theta_{LVC})$ values and the $\mathcal{D}$ statistic is computed for each subsample with observational uncertainties propagated. The error bars in each bin represent the $1\sigma$ value of $\mathcal{D}$ from a bootstrap resampling of the data $10^3$ times per bin.
}
\label{fig:NcUp15_P-and-D-vs-Dtheta}
\end{center}
\end{figure}

To test this simple scenario, we consider all lines of sight showing a sufficient degree of complexity in terms of number of clouds, namely $\mathcal{N}_c > 1.5$.  We bin the sky pixels according to their $\Delta(\theta_{LVC},\theta_{IVC})$ values as measured from \hi orientation data in the scheme of Implementation I.
Then, for each bin, we examine the distribution of $p_{353}$ and compute the $\mathcal{D}$ statistic.
As expected, for increasing $\Delta(\theta_{LVC},\theta_{IVC})$, we observe a small but systematic decrease of degree of polarization and a clear rise of $\mathcal{D}$ values.
The latter is shown in Fig.~\ref{fig:NcUp15_P-and-D-vs-Dtheta} for the PR3 polarization maps from which the \texttt{smica} CMB has been subtracted.
We obtain similar conclusions when we use other combinations of polarization maps and removed CMB estimates, as well as when we consider the \hi orientation as in Implementation II of the selection of \texttt{target} pixels (i.e., at the peak of the two dominant clouds) rather than the scheme used in Implementation I.

In the simple two-cloud model of \citet{Tas2015}, LOS decorrelation is expected to be more pronounced towards lines of sight where the magnetic fields of the clouds form an angle of $60^\circ$ or more. As noted by these authors, smaller angle differences can also result in LOS decorrelation, but at a lower level. LOS decorrelation is therefore not expected to abruptly appear at some large misalignment angle, but should qualitatively match the observed smooth trend in Fig. \ref{fig:NcUp15_P-and-D-vs-Dtheta}. A more quantitative comparison of this observation with analytic models should take into account a number of factors. First, in general, lines of sight might be composed of more than two dust clouds that contribute to the polarized signal. Second, changes in the spectral index of the dust SED (and not simply the dust temperature, as assumed in the \citet{Tas2015} model) can alter the frequency dependence of the dust emission EVPA for a given misalignment angle. Finally, the difference between \hi filament orientation and the plane-of-the-sky (POS) magnetic field orientation shows an intrinsic astrophysical scatter, which should also be taken into account as an extra source of uncertainty. Such detailed comparisons with models will require further work beyond that presented in this paper.

We note that in our analysis we have {\em not} optimized our cutoff in $\Delta(\theta_{LVC},\theta_{IVC})$ for the selection of our \texttt{target} pixels; rather, we adopted $60^\circ$ based on our {\it a priori} physical expectations. Had we decreased the cutoff to $\Delta(\theta_{LVC},\theta_{IVC}) \geq 45^\circ$, the size of the \texttt{target} sample, and hence the significance with which we have detected LOS frequency decorrelation, would have increased, as subsequent analysis confirms.

\subsection{A case study using starlight polarization}
\label{subsec:appetizer}
In this paper we have used \hi morphology as an indirect probe of the direction of magnetic fields in individual clouds. 
Starlight polarization, induced by the same dust grains that produce polarized emission, is a more direct probe of the dust
polarization position angle.
Currently available starlight polarization measurements are sparse, but large-scale starlight polarization surveys
like \textsc{Pasiphae} (\citealt{PASIPHAEwhite}) are planned for the near future.
Nevertheless, data do exist in a small sky patch that we can use for a proof-of-principle analysis using starlight polarization instead of \hi data.

\begin{figure}
\begin{center}
\begin{tabular}{c}
\includegraphics[trim={2.cm .4cm 1.25cm 1.2cm},clip,height=.68\columnwidth]{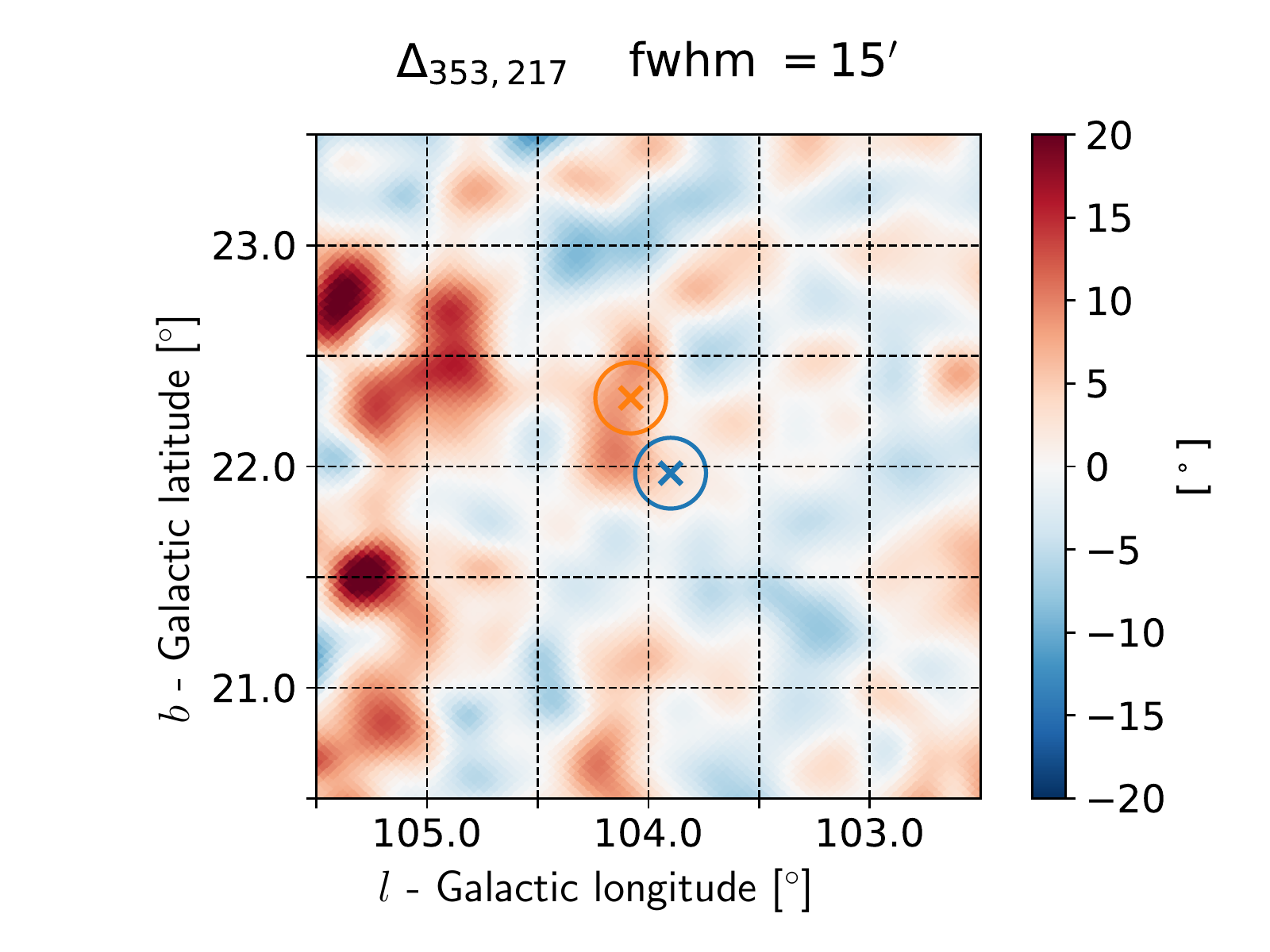}
\rotatebox{90}{ \hspace{2.cm} $\Delta_s(\psi_{353},\psi_{217})$ [$^\circ$]}
\\
\includegraphics[trim={.2cm 1.2cm .2cm 1.2cm},clip,height=.68\columnwidth]{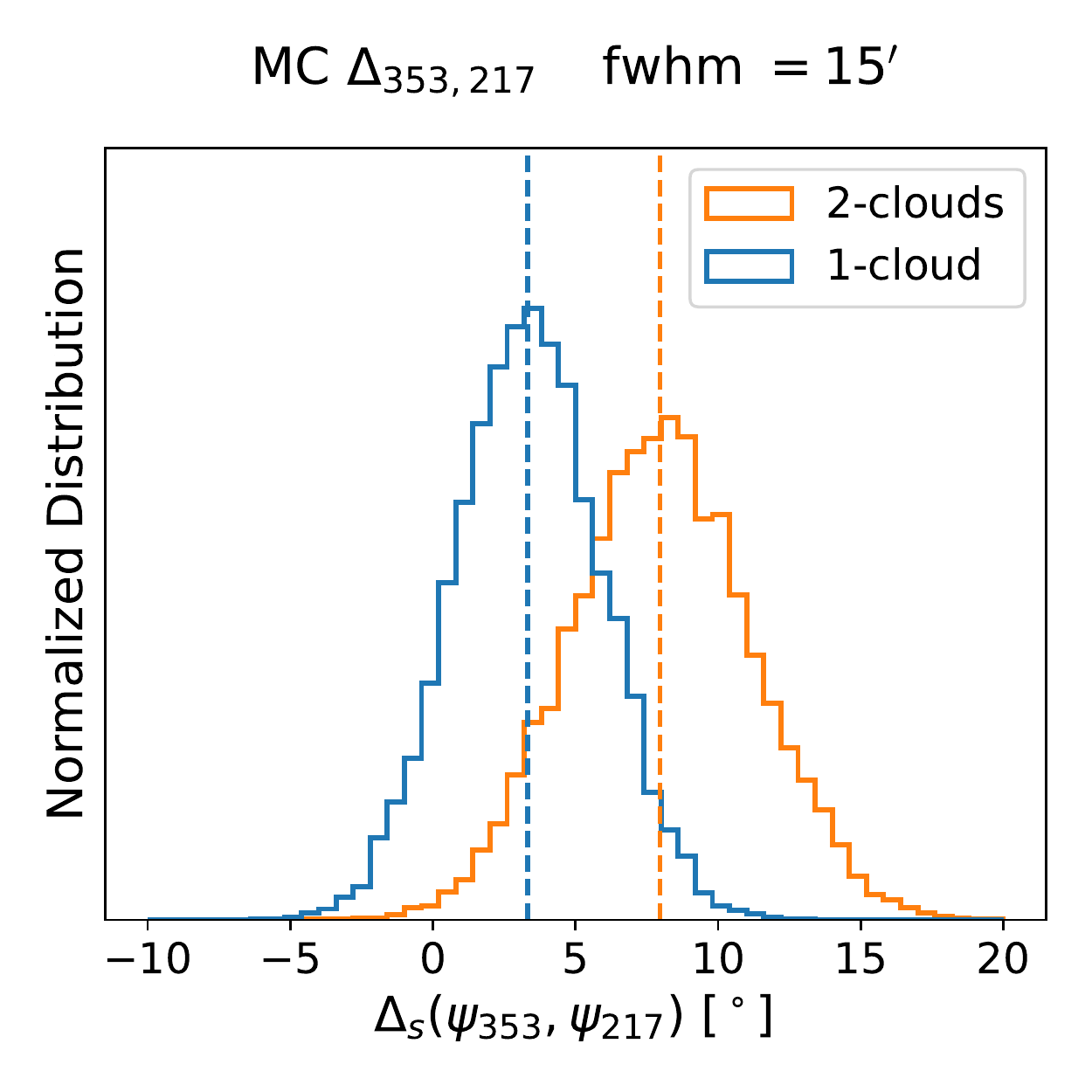}\\
$\Delta_s(\psi_{353},\psi_{217})$ [$^\circ$]
\end{tabular}
\caption{The case of tomography region of \cite{Pan2019}. Top panel: Map of EVPA differences computed from 353 and 217 GHz polarization maps from {\it Planck}. The 2-cloud and 1-cloud sight lines are marked respectively by orange and blue crosses at North-East and South-West of the map center. The circles have 16\arcmin ~radius and mark the beams within which starlight polarization data have been taken and studied by \cite{Pan2019}.
Bottom panel: Histograms of EVPA differences computed through 10,000 MC simulations to propagate observational uncertainties on ($Q_\nu,\,U_\nu$). The 2-cloud LOS histogram is shown in orange, the 1-cloud LOS in blue. The vertical lines, with corresponding colors, show the EVPA differences from the data.
}
\label{fig:TomoRegion}
\end{center}
\end{figure}

\cite{Pan2019} used starlight polarization data from the RoboPol polarimeter \citep{RobopolInstrument} to study a sky region where several Galactic dust components are present along the LOS.
Based on these stellar polarization data the authors inferred the number of dust clouds and the POS orientation of the magnetic field permeating those for two nearby observing beams of 16\arcmin ~radius. The latter two were pre-selected based on \hi data to likely harbor two dust clouds (2-cloud LOS) and one dust cloud (1-cloud LOS).

The authors demonstrated that the two clouds exhibit significant differences in terms of column density and polarization properties, and that their mean POS magnetic field orientations differ by about $60^\circ$. In principle, the different SEDs in those significantly misaligned clouds could lead to a measurable effect of LOS frequency decorrelation in {\it Planck} data towards the 2-cloud LOS. However, if the effect is weak it could be hidden in the noise, as suggested in Sect.~6.3 of \cite{Pan2019} based on a set of polarization maps from the second {\it Planck} data release.

Here, we investigate further the polarization data for those particular lines of sight. We retrieve the polarized emission at 217 and 353 GHz measured by {\it Planck} towards the sky region of interest (see Sect.~\ref{sec:data}) smoothed to a 16\arcmin ~FWHM beam, and we compute the signed difference of EVPA in each pixel (see Eq.~\ref{eq:Delta_s} in Sect.~\ref{sec:analysis}).
We thus obtain the EVPA-difference map presented in Fig.~\ref{fig:TomoRegion} (top) where we highlight the two LOS studied in \cite{Pan2019}. Interestingly, the 2-cloud region at the center, which is known to feature a complex magnetized ISM structure with at least two dust components (\citealt{Pan2019}, \citealt{Cla2019}), displays a higher EVPA difference than the nearby 1-cloud region.

We quantify the level of uncertainty of the EVPA difference induced by the observational uncertainties on the Stokes $Q_\nu$ and $U_\nu$ ($\nu= \left\lbrace 217,\, 353 \right\rbrace$) through MC simulations (see Sect.~\ref{sec:data} for details). For each MC draw, we compute the EVPA difference and build the histograms shown in the right panel of Fig.~\ref{fig:TomoRegion}. Even when accounting for {\it Planck} noise, the 2-cloud LOS deviates significantly from zero EVPA difference in the two frequencies, suggesting a LOS frequency decorrelation of the polarization data. In contrast, the distribution corresponding to the 1-cloud LOS is compatible with zero EVPA difference in the two frequencies (no LOS frequency decorrelation). Although less significant, the offset from zero of the EVPA difference for the 2-cloud LOS survives the subtraction of the CMB estimates. This is reported in Table~\ref{tab:TR-results}.
\begin{table}
\caption{EVPA frequency differences in degrees for the 2-cloud and 1-cloud LOSs from PR3 frequency maps and with subtraction of \texttt{commander} and \texttt{smica} CMB estimates.
The means and 1$\sigma$ intervals are computed through 10,000 MC simulations to propagate the observational uncertainties. 68\% of the draws fall within the quoted uncertainty about the mean.
}
\label{tab:TR-results}
\centering
\begin{tabular}{l c c c}
\hline\hline
\\[-1.5ex]
CMB Removal  &   {1-cloud LOS}  &  {2-cloud LOS} \\
\\[-1.5ex]
\hline \\[-1.5ex]
None  & $3.30\pm2.54$ [$^\circ$] & $7.98\pm3.10$ [$^\circ$]   \\
\texttt{commander} & $2.91\pm2.47$ [$^\circ$] & $6.44\pm3.27$ [$^\circ$]  \\
\texttt{smica} & $3.06\pm2.48$ [$^\circ$]  & $7.33\pm3.24$ [$^\circ$]  \\
\\[-1.5ex]
\hline
\end{tabular}
\end{table}
This tentative result demonstrates how starlight polarization data can be used to identify sky pixels that experience LOS-induced frequency decorrelation.

\section{Estimation of required SED variation}
\label{sec:SEDVar}
Frequency decorrelation of dust emission is, ultimately, the result of spatial variations of the dust SED. The detection of rotation of the dust polarization angle between frequencies is evidence for variation of the dust SED along the line of sight. We can therefore use the observed magnitude of this effect to estimate the intrinsic variability of the dust SED.

Let us divide the line of sight into $N$ clouds such that the $i$-th cloud has column density $N_{\rm HI}^i$. Then the observed Stokes parameters of the polarized dust emission at a frequency $\nu$ are given by \citep[e.g.,][]{Hen2019}:

\begin{align}
Q_\nu &= \sum_i m_p N_{\rm HI}^i \delta_{\rm DG}^i f^i \kappa_\nu^i B_\nu\left(T_d^i\right)\cos^2\gamma_i\cos\left(2\psi_i\right) \label{eq:q_model} \\
U_\nu &= \sum_i m_p N_{\rm HI}^i \delta_{\rm DG}^i f^i \kappa_\nu^i B_\nu\left(T_d^i\right)\cos^2\gamma_i\sin\left(2\psi_i\right) \; , \label{eq:u_model}
\end{align}
where $f^i$, $\delta_{\rm DG}^i$, $\kappa_\nu^i$, $T_d^i$, $\gamma_i$, and $\psi_i$ are the alignment fraction, dust-to-gas mass ratio, polarized opacity at frequency $\nu$, dust temperature, angle between the magnetic field and the plane of the sky, and polarization angle of the $i$th cloud, respectively, and $m_p$ is the proton mass. When there are multiple clouds along the line of sight, Eqs.~\ref{eq:q_model} and~\ref{eq:u_model} make clear that the ratio $U_\nu/Q_\nu$, and thus the polarization angle $\psi_\nu$, is generally not constant with frequency.

For a single cloud, the ratios of $Q_\nu$ and $U_\nu$ at 217 and 353~GHz are given by

\begin{equation}
\label{eq:ratio}
    \left(\frac{Q_{217}}{Q_{353}}\right)_i = \left(\frac{U_{217}}{U_{353}}\right)_i = \frac{B_{217}\left(T_d^i\right)\kappa_{217}^i}{B_{353}\left(T_d^i\right)\kappa_{353}^i} \; . \\
\end{equation}
If dust everywhere had the same temperature and same opacity law, then this ratio would be constant across the sky and $\psi_\nu$ would be constant with frequency. Since this is inconsistent with what is observed, let us assume that this quantity has a mean value $\alpha$ and that cloud-to-cloud variations are described by a parameter $\rho$ having mean zero, i.e.,

\begin{equation}
    \label{eq:rho}
    \left(\frac{Q_{217}}{Q_{353}}\right)_i = \left(\frac{U_{217}}{U_{353}}\right)_i
    \equiv \alpha\left(1 + \rho_i\right) \; .
\end{equation}
A modified blackbody having $T_d = 19.6$\,K and $\beta = 1.55$, typical parameters for high-latitude dust \citep{PlaXI2018}, has $\alpha = 0.21$, though our analysis is not sensitive to the value of $\alpha$. $\sigma_\rho$ quantifies the intrinsic variation in the dust SED between 217 and 353~GHz, regardless of whether those variations arise from temperature, composition, or other effects. Modeling $\rho$ as Gaussian distributed with mean zero and variance $\sigma_\rho^2$, we seek the value of $\sigma_\rho$ that can account for the enhanced dispersion of polarization angles on multi-cloud sightlines (Fig.~\ref{fig:DPPAs}).

To estimate the effect of $\sigma_\rho$ on the dispersion in polarization angles, we use the \hi maps to constrain both the line-of-sight distribution of clouds and their relative orientations.
To simplify the analysis we consider the data from our Implementation~2 (Sect.~\ref{sec:analysis}). For each sightline we thus consider only the two dominant clouds (in \hi column density) as identified by \citet{Pan2020} and create maps of per-cloud Stokes parameter by integrating the \HI-based $Q$ and $U$ maps of (\citealt{Cla2019}) in the velocity range within $v_0 \pm \sigma_0$, where $v_0$ is the cloud centroid velocity and $\sigma_0$ is the second moment of its spectrum (see Sect.~\ref{sec:data}). Then, we estimate $\psi_{353}$ on each sightline as:
\begin{equation}
    \hat{\psi}_{353} = \frac{1}{2} \arctan\left(\frac{U_{\rm HI}^1\cos^2\gamma_1 + U_{\rm HI}^2\cos^2\gamma_2}{Q_{\rm HI}^1\cos^2\gamma_1 + Q_{\rm HI}^2\cos^2\gamma_2}\right) \; ,
\end{equation}
where $Q_{\rm HI}$ and $U_{\rm HI}$ are given by Eqs.~\ref{eq:Q_HI} and~\ref{eq:U_HI}, respectively, and the superscripts denote integration over clouds 1 and 2.
The angles between the magnetic field and the plane of the sky $\gamma_1$ and $\gamma_2$ are unknown, and so we draw $\sin\gamma$ uniformly from the interval $[-1,1]$ for each; $\gamma = 0$ when the magnetic field is in the plane of the sky. This equation does not explicitly model variations in the 353~GHz dust emissivity per H atom, although marginalizing over different values of $\gamma_1$ and $\gamma_2$ achieves a similar effect numerically. Rather, since we are interested only in the variability of the polarized dust SED between 353 and 217~GHz, we model such effects through the $\rho_1$ and $\rho_2$ parameters when computing the 217~GHz polarization angle only.

Using Eq.~\ref{eq:rho}, $\psi_{217}$ on each sightline can be modeled as
\begin{equation}
    \hat{\psi}_{217} = \frac{1}{2} \arctan\left(\frac{U_{\rm HI}^1\left(1 + \rho_1\right)\cos^2\gamma_1 + U_{\rm HI}^2\left(1 + \rho_2\right)\cos^2\gamma_2}{Q_{\rm HI}^1\left(1 + \rho_1\right)\cos^2\gamma_1 + Q_{\rm HI}^2\left(1 + \rho_2\right)\cos^2\gamma_2}\right) \; .
\end{equation}
On each sightline, $\rho_1$ and $\rho_2$ are drawn from a Gaussian distribution of mean zero and variance $\sigma_\rho^2$. Then for each sightline we can compute $\Delta_s\left(\hat{\psi}_{353},\hat{\psi}_{217}\right)$ (Eq.~\ref{eq:Delta_s}) and finally the dispersion $\mathcal{D}$ (Eq.~\ref{eq:D}) over all \texttt{target2} sightlines as a function of $\sigma_\rho$.

\begin{figure}
    \centering
        \includegraphics[trim={.2cm .2cm .0cm .0cm},clip,height=6.4cm]{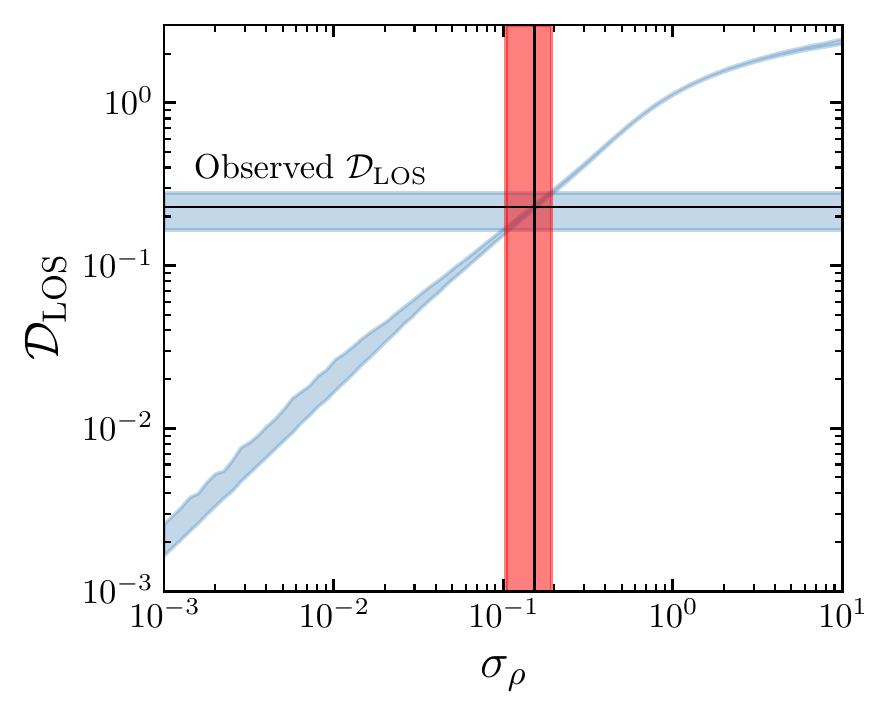}
    \caption{The dispersion $\mathcal{D}_{\rm LOS}$ (Eq. \ref{eq:dlos}) resulting solely from variations in the dust SEDs between two clouds along the line of sight in the \texttt{target2} sample. We quantify the level of SED variation by the parameter $\sigma_\rho$ (Eq.~\ref{eq:rho}), finding that $\sigma_\rho = 0.15$ can account for the excess dispersion in the \texttt{target} sample. Thus, we estimate that the ratio of 353 to 217~GHz polarized intensity is varying at roughly the 15\% level from cloud to cloud. The blue shaded regions indicate the observed range of $\mathcal{D}$ estimated in Sect.~\ref{sec:analysis} and the $\pm1\sigma$ confidence interval from 1000 realizations of $\gamma_1$, $\gamma_2$, $\rho_1$, and $\rho_2$ in each pixel. The red shaded region is the resultant constraint on $\sigma_\rho$.
    } \label{fig:sigma_rho}
\end{figure}

If the difference in dispersion between the \texttt{target} and \texttt{control} samples is attributed entirely to varying dust SEDs, then we can estimate

\begin{equation}
    \mathcal{D}_{\rm LOS} \simeq \sqrt{\mathcal{D}\left({\rm target}\right)^2 - \mathcal{D}\left({\rm control}\right)^2} = 0.23^{+0.05}_{-0.06}
    \label{eq:dlos}
\end{equation}
from the resampling analysis presented in Sect.~\ref{sec:analysis} and Table~\ref{tab:SL_F3}. This range is indicated by the horizontal band in Fig.~\ref{fig:sigma_rho}.

The $\pm1\sigma$ range of $\mathcal{D}$ over 1000 simulations for each value of $\sigma_\rho$ is presented in Fig.~\ref{fig:sigma_rho}. We see that $\sigma_\rho = 0.15$ matches the observed enhancement in dispersion between the \texttt{target} and \texttt{control} samples. This is consistent with the dust SEDs varying in the ratio of 217 to 353~GHz polarized intensity at the level of 15\% from cloud to cloud over the region analyzed. As we model contributions from only the two most dominant clouds on each sightline, we may be slightly overestimating the true dispersion.

\section{Discussion}
\label{sec:discussion}
In this paper we report on the detection of the effect of LOS-induced frequency decorrelation -- the combined effect of varying dust SEDs and magnetic field orientations along the LOS -- in {\it Planck} polarization data.
This detection was made possible by the use of \hi datasets, which allowed us to construct our \texttt{target} and \texttt{control} samples  {\it a priori}, in an astrophysically motivated way.
The consistency of the results between our two implementations and between the different sets of polarization maps and CMB estimates reinforces our confidence that our finding is robust.
Our analysis has additionally shown that the significance of the effect becomes higher when we use maps cleaned from residual systematics that were present in {\it Planck} PR3 polarization maps.

We emphasize that we have not in any way optimized our analysis choices to maximize the significance with which the effect is detected. Rather, whenever a choice had to be made, we made it based on astrophysical arguments. There are several examples where different choices in our analysis would have, in fact, {\em increased} the significance of the detection of the effect (decreased the p-value of Null Hypotheses I and II).
These include:\\ (a) Definition of \texttt{target}: defining \texttt{target} as the union of \texttt{target1} and \texttt{target2} increases the significance. \\
(b) Cutoff in \hi orientation misalignment: changing the misalignment requirement for inclusion in \texttt{target} from $\geq 60^\circ$ to $\geq 45^\circ$ increases the significance. \\
(c) Localization: restricting our analysis to the northern hemisphere increases the significance.

This work finds evidence for LOS frequency decorrelation, and does not directly address the question of decorrelation in the dust power spectra.
Our findings show that frequency decorrelation of the dust polarization signal is not an effect that is uniform throughout the sky since the change in polarization pattern is more severe for sightlines that pass through more convoluted magnetized ISM, and that those particular sightlines are distributed unevenly on the sky (see the bottom map in Fig.~\ref{fig:dataMap-orth_Nc15}). This may have implications in power-spectrum--based estimates as future work will clarify.

From a CMB perspective, it would be interesting to estimate the level of LOS-induced frequency decorrelation using cross-power spectra (as in, e.g., \citealt{PlaXXX2016}; \citealt{PlaXI2018}) on maps that have been corrected for residual systematics (\citealt{Del2019}; \citealt{PlaLVII2020}) and in sky regions that are dominated by pixels comprising our \texttt{target} samples. Such an analysis is beyond the scope of this paper.

Our Implementation I of the \texttt{target} pixel selection focused on the distinction between LVCs and IVCs, based on the physical expectation that IVCs might feature different dust SEDs than LVCs, due to differences in temperature and/or dust grain properties (\citealt{PlaXXIV2011}; \citealt{PlaXI2014}). Our Implementation II imposed no such constraint on the velocities of the identified distinct peaks in \hi emission. This enables us to use \texttt{target2} to {\it a posteriori} test whether IVC-LVC cloud pairs exhibit a stronger LOS frequency decorrelation effect than LVC-LVC pairs. 

Concentrating on the two dominant clouds (the ones corresponding to the two highest-\HI-column-density components), we split pixels in \texttt{target2} in two groups: pixels
where both dominant clouds have a velocity centroid in the LVC range (513 pixels), and pixels dominated by LVC-IVC pairs (5242 pixels), with LVC and IVC ranges defined as in Sect.~\ref{subsec:pixelsamples}. 
We infer the relative strength of LOS frequency decorrelation in the two groups through a uniform, unweighted resampling analysis of each subset of pixels, with $N_{\rm{Boot}} = 500$.
The $\mathcal{D}$ values obtained for PR3 polarization maps and \texttt{smica} CMB subtraction are
$\mathcal{D}_{\rm{IVC-LVC}}$ = $ 1.05 \pm 0.04$ and  $\mathcal{D}_{\rm{LVC-LVC}}$ = $ 1.13 \pm 0.05$.
The $\mathcal{D}$ statistic is thus found to be higher for LVC-LVC pairs than for IVC-LVC pairs, although the two values are consistent within sampling uncertainties. The same trend is observed for all combinations of polarization maps and subtracted CMB estimate. As a result, it appears that the LOS frequency decorrelation induced by dust clouds does not {\it only} involve lines of sight passing through IVCs.
And, on the contrary, significantly misaligned LVCs may be a substantial source of LOS frequency decorrelation.

To improve CMB dust polarization foreground modeling and subtraction, accounting for LOS frequency decorrelation, observables that can provide insight on the 3D structure of the magnetized ISM will play a critical role. Such observables include \hi data (as we have done in this paper) and starlight polarization. Starlight polarization originates in dichroic absorption by the same dust grains that produce polarized emission, and thus traces the same physical processes of grain alignment with the magnetic field, but for the line of sight between observer and star.
Large-scale starlight polarization surveys like \textsc{Pasiphae} (\citealt{PASIPHAEwhite}) will thus soon provide an independent, direct probe of dust grain orientations in individual clouds.

\section{Conclusions}
\label{sec:conclusions}

That the SEDs of the dust clouds vary to some extent between different parts of the Galaxy is certain.  That there are in general multiple dust clouds along a large fraction of lines of sight is certain. That the magnetic field of the Galaxy is not uniform and may vary along the LOS is certain. 
Consequently, decorrelation between polarized dust emission at different frequencies,  both in the plane of the sky and along the LOS, must be present to some extent. 
The relevant question is whether the magnitude of this frequency decorrelation effect is high enough to be detected by an instrument of given specifications.

In this work, we pursue a new approach that specifically targets LOS frequency decorrelation. 
Physically, we expect that LOS frequency decorrelation does not occur at a uniform level throughout the sky, but rather should be more severe where the orientations of the magnetic field permeating different dust clouds superposed along the LOS are strongly misaligned. Therefore, we used \hi velocity and orientation data to select pixels that are most likely to exhibit significant LOS frequency decorrelation induced by multiple dust SED components.
Each of these target sightlines has an \hi emission structure consistent with multiple LOS clouds with misaligned magnetic fields. We compare these to a control sample of sightlines that contain a single \hi cloud.
The use of \hi allows us to distinguish these two sets of pixels using data that are entirely independent of polarization measurements. 
The pixels that maximize the likelihood of showing a LOS frequency decorrelation signal are highly non-evenly distributed on the sky.

We quantify LOS decorrelation using the dispersion of inter-frequency EVPA differences. We find that this dispersion is larger for our target sample than for the control sample in \textit{Planck} data. We detect the LOS frequency decorrelation effect at a level above the {\it Planck} noise (see Fig.~\ref{fig:D_wBDPPAs}).
We have confirmed that our finding is robust 
to inhomogeneous data noise level, residual systematics, CMB contamination, or the specifics of sky pixel selection.
We found that trends in polarization data follow closely the phenomenology expected from the simplest modeling of the effect (Fig.~\ref{fig:NcUp15_P-and-D-vs-Dtheta}).
Additionally, relying on a model-independent approach, we estimated that an intrinsic variability of the dust SED of $\sim 15\%$ can lead to the observed magnitude of the effect that we measured from polarization maps at 353 and 217 GHz (see Fig.~\ref{fig:sigma_rho}).
Finally, we demonstrated that LOS superposition of both LVC-LVC and LVC-IVC pairs of clouds contributes to the signal detection.

\medskip

In this study we have presented the first detection of LOS frequency decorrelation in the {\it Planck} data. This detection was made possible thanks to the use of ancillary datasets, \hi emission data and starlight polarization data, that allow us to identify sky regions that are potentially most susceptible to this effect.

\medskip


\begin{acknowledgements}
We thank Vincent Guillet and Aris Tritsis for insightful discussions.
We thank our anonymous referee for her/his report.
This work has received funding from the European Research Council (ERC) under the European Union's Horizon 2020 research and innovation programme under grant agreement Nos. 771282, 772253, and 819478.
G. V. P. acknowledges support by NASA through the NASA Hubble Fellowship grant  HST-HF2-51444.001-A  awarded  by  the  Space Telescope  Science  Institute,  which  is  operated  by  the Association of Universities for Research in Astronomy, Incorporated, under NASA contract NAS5-26555. S. E. C. acknowledges support by the Friends of the Institute for Advanced Study Membership.
V. P. acknowledges support from the Foundation of Research and Technology - Hellas Synergy Grants Program through project MagMASim, jointly implemented by the Institute of Astrophysics and the Institute of Applied and Computational Mathematics and by the Hellenic Foundation for Research and Innovation (H.F.R.I.) under the “First Call for H.F.R.I. Research Projects to support Faculty members and Researchers and the procurement of high-cost research equipment grant” (Project 1552 CIRCE).
We acknowledge the use of data from the Planck/ESA mission, downloaded from the Planck Legacy Archive, and of the Legacy Archive for Microwave Background Data Analysis (LAMBDA). Support for LAMBDA is provided by the NASA Office of Space Science.
Some of the results in this paper have been derived using the HEALPix (G{\'o}rski et al. 2005) package. This work is partially based on publicly released data from the HI4PI survey which combines the Effelsberg–Bonn HI Survey (EBHIS) in the northern hemisphere with the Galactic All-Sky Survey (GASS) in the southern hemisphere.
\end{acknowledgements}

%
\bibliographystyle{aa} 
\bibliography{myDecorrBib} 
%
\end{document}